\documentclass[12pt]{article}
\pdfoutput=1
\usepackage[T1]{fontenc}
\usepackage{fullpage}
\usepackage{amsmath,amsfonts,amssymb,graphicx,color}
\DeclareMathOperator{\Tr}{Tr}
\usepackage{authblk}

\usepackage{amsthm}

\usepackage{listings}
\usepackage{mathrsfs}

\usepackage[unicode=true,bookmarks=true,
bookmarksnumbered=false,bookmarksopen=false,
breaklinks=false,
pdfborder={0 0 1},
backref=false,colorlinks=true]{hyperref}
\usepackage{xcolor}
\usepackage{verbatim}
\usepackage{cite}
\usepackage[hang,flushmargin]{footmisc} 

\hypersetup{pdftitle={Traversable Wormholes as Quantum Channels},
 pdfauthor={Ning Bao, Aidan Chatwin-Davies, Jason Pollack, and Grant N. Remmen},
 citecolor=black,linkcolor=black,urlcolor=black}

\newcommand{\ket}[1]{\left| #1 \right\rangle}
\newcommand{\bra}[1]{\left\langle #1 \right |}
\def\({\left(}
\def\){\right)}
\def\[{\left[}
\def\]{\right]}

\newenvironment{emergency}[1]{%
  \par
  \setlength{\emergencystretch}{#1}%
}{%
  \par
}

\newcommand{\be}{\begin{equation}}
\newcommand{\ee}{\end{equation}}
\newcommand{\Eq}[1]{Eq.~(\ref{#1})}
\newcommand{\Eqs}[2]{Eqs.~(\ref{#1}) and (\ref{#2})}
\newcommand{\Sec}[1]{Sec.~\ref{#1}}
\newcommand{\Secs}[2]{Secs.~\ref{#1} and \ref{#2}}
\newcommand{\Fig}[1]{Fig.~\ref{#1}}

\newcommand{\Ref}[1]{Ref.~\cite{#1}}

\newcommand{\mrm}[1]{\mathrm{#1}}

\newcommand{\Hil}{\mathcal{H}}
\newcommand{\oh}{\mathcal{O}}
\newcommand{\N}{\mathcal{N}}
\newcommand{\TFD}{\mathrm{TFD}}
\DeclareMathOperator{\myRe}{Re}

\newtheorem{thm}{Theorem}[section]

\newtheorem{defn}[thm]{Definition}

\newtheorem{example}[thm]{Example}

\newcommand{\ketbra}[2]{|#1\rangle \langle #2 |}

\newcommand{\tket}[1]{| #1 \rangle}
\newcommand{\tbra}[1]{\langle #1 |}

\begin{document}

\interfootnotelinepenalty=10000

\hfill CALT-TH-2018-032
\hfill

\vspace{2cm}
\thispagestyle{empty}
\begin{center}
{\LARGE \bf
Traversable Wormholes as Quantum Channels: \\Exploring CFT Entanglement Structure and Channel Capacity in Holography
}\\
\bigskip\vspace{1cm}{
{\large Ning Bao,${}^a$ Aidan Chatwin-Davies,${}^{b,c}$ Jason Pollack,${}^d$ and Grant N. Remmen${}^a$}
} \\[7mm]
 {\it ${}^a$Center for Theoretical Physics and Department of Physics \\
     University of California, Berkeley, CA 94720, USA and \\
     Lawrence Berkeley National Laboratory, Berkeley, CA 94720, USA \\[1.5mm]
${}^b$Walter Burke Institute for Theoretical Physics \\
    California Institute of Technology, Pasadena, CA 91125, USA\\[1.5mm]
         ${}^c$KU Leuven, Institute for Theoretical Physics\\Celestijnenlaan 200D 
B-3001 Leuven, Belgium\\[1.5mm]
 ${}^d$Department of Physics and Astronomy\\University of British Columbia, Vancouver, BC, V6T 1Z1, Canada} \let\thefootnote\relax\footnote{\noindent e-mail:\\\url{ningbao75@gmail.com}, \url{aechatwi@gmail.com}, \url{jpollack@phas.ubc.ca}, \url{grant.remmen@berkeley.edu}} \\
 \end{center}
\bigskip
\centerline{\large\bf Abstract}

\begin{quote} \small

We interpret the traversable wormhole in AdS/CFT in the context of quantum information theory. In particular, we investigate its properties as both a quantum channel and entanglement witness. We define protocols that allow either the bounding of the channel's entanglement capacity or the determination of aspects of the entanglement structure between the two boundary CFTs. Such protocols and connections allow for the use of quantum channel techniques in the study of gravitational physics and vice versa. More generally, our results suggest a purely quantum information-theoretic criterion for recognizing when the product of two boundary theories has a classical bulk interpretation.
\end{quote} 

\setcounter{footnote}{0}

\newpage
\tableofcontents
\newpage

\section{Introduction}
One of the most intriguing results of the recent exploration of the relationship between quantum information and quantum gravity in holography \cite{tHooft:1993dmi,Susskind:1994vu,Bousso:2002ju}, in the particular context of the AdS/CFT correspondence \cite{Maldacena:1997re,Gubser:1998bc,Witten:1998qj,Aharony:1999ti}, is the realization that black holes connected by a wormhole are highly quantum-mechanically entangled with each other \cite{Israel:1976ur,Maldacena:2001kr,Maldacena:2013xja}. Thought experiments suggested by this realization, in which connected black holes are treated as entangled quantum states \cite{Jensen:2013ora,Sonner:2013mba,Bao:2015nqa, Bao:2015nca, Remmen:2016wax}, have elucidated connections between general-relativistic results for the wormhole geometry and quantum-mechanical results concerning entangled states. Such thought experiments can often be viewed as probing the classical, many-qubit limit of the proposed ER/EPR conjecture \cite{Maldacena:2013xja}, which relates quantum entanglement and wormholes more generally.

Typically, the imposition of causality and energy conditions prevents anything from traveling from one side of the Einstein-Rosen bridge to the other \cite{Morris:1988tu}. However, it has been shown that wormholes can be rendered traversable in AdS/CFT \cite{Gao:2016bin, Maldacena:2017axo} via the insertion of a double-trace deformation on the boundary CFTs: in effect, as we review in \Sec{sec:AdSCFT}, a bilocal operator coupling the two CFTs introduces a negative null energy shock wave in the bulk and hence allows causal paths through the wormhole. In previous work \cite{Bao:2015nca}, it was shown in the classical holographic limit that the inability to differentiate with perfect confidence between a pair of black holes either connected by a classical interior Einstein-Rosen bridge geometry or not is dual to the quantum-mechanical fact that entanglement is not a linear observable. As an exercise, in \Sec{sec:observable} we revisit this result in the traversable wormhole context and show it continues to hold, as it must given that the boundary is still described by a good quantum-mechanical theory.

More generally, however, we would like to examine the broader implications of the traversable wormhole construction in the context of quantum information theory. We know that when a wormhole is rendered traversable we can use it to send (some) signals between two regions of spacetime or, equivalently, between subsystems of the two dual CFTs.
A natural question is whether the propagation of such signals through the would-be traversable wormhole region can function as a reliable entanglement witness\footnote{An entanglement witness is an operator that is capable of distinguishing certain patterns of entanglement from separable states. For a more detailed definition, we refer the reader to \Sec{sec:witness}.} for the quantum system of the two entangled black holes.

Furthermore, the successful transmission of such signals manifestly results in the transfer of information between the two regions (or the two CFTs).
While the overall evolution of the two CFTs is jointly unitary, the transport of quantum information from one CFT to another is a process that, since it concerns subsystems, need not be unitary (and, in fact, the transport of qubits is generically nonunitary in everyday laboratory situations where they cannot be totally isolated from their environments). The process of sending information through a wormhole is thus better described in the language of a quantum channel.\footnote{A quantum channel provides a general formalism for describing the transmission of quantum (and classical) information, not necessarily unitarily, in a potentially noisy system; see \Sec{sec:channels}.}
Given this description, we would like to better understand the role that the traversable wormhole is playing as a quantum communication channel between the two CFTs and whether its bulk properties translate into any nontrivial properties of the relevant subclass of quantum channels. At the same time, we can also use signal propagation via such a quantum channel to learn about the structure of the wormhole itself. 

In this paper, we formalize both of these notions, constructing algorithms that one would follow to utilize the traversable wormhole either as a quantum channel for sending information or as an entanglement witness to probe the spacetime geometry and its dual entanglement structure.
After some preliminaries, we define (in \Sec{sub:channel_subregions}) the quantum channel that evolves excitations initially localized near a part of one boundary subregion to excitations near the other boundary.
Because this is a channel between infinite-dimensional Hilbert spaces, we are next motivated to consider (in \Sec{sub:channel_code}) a coarse-grained mapping between finite-dimensional code subspaces, following \Ref{Almheiri:2014lwa}. 
The construction of this channel allows us to make contact with the quantum information literature on finite-dimensional channels, and (in \Sec{sec:quantum_channel_capacity_wormhole}) we combine these results with the gravitational description of the channel to place a bound on its entanglement capacity and describe a protocol that saturates this bound.
While this protocol maximizes the number of qubits that can be sent through the wormhole while it remains traversable, it is not optimized to measure properties of the deformation itself. 
We thus consider (in \Sec{sec:TWEW}) additional protocols that better exploit the nature of the channel as an entanglement witness for the bulk spacetime geometry.

Throughout this paper, we will work in a semiclassical approximation, where we can take the spacetime geometry of the wormhole to be well described by Einstein's equations, corresponding under AdS/CFT to two entangled black holes in the large-$N$ limit. In particular, we will not use the assumptions of ER/EPR \cite{Maldacena:2013xja}, since we do not need to ascribe any geometric notion to single Bell pairs or small numbers of qubits.

The organization of this paper is as follows. In \Sec{sec:channels}, we discuss some formalism for quantum channels. In \Sec{sec:AdSCFT}, we review the traversable wormhole construction in AdS/CFT. In \Sec{sec:observable}, we comment on the implications of wormhole traversablity for the observability of entangled states. In \Sec{sec:TWQC}, we put these concepts together for a rigorous definition of traversable wormholes as a specific class of quantum channels. In \Sec{sec:witness}, we discuss entanglement witnesses in quantum information theory. Finally, in \Sec{sec:TWEW} we construct a setup in which traversable wormholes can serve as partial entanglement witnesses for the class of quantum states of pairs of black holes with unknown mutual entanglement structure. We conclude in \Sec{sec:discussion} with some final discussion and thoughts on future work.

\section{Review of Quantum Channels}\label{sec:channels}

We begin with a brief review of quantum channels and the associated technology relevant to the analysis of traversable wormholes.
A more complete treatment of the subject can be found in Refs.~\cite{bib:nc,Preskill:2016htv}.

\subsection{Channel basics}

A quantum channel generalizes the notion of unitary evolution in quantum mechanics to include the possibility of dissipative evolution.
Quantum channels map density matrices onto density matrices, but information need not be preserved by this mapping.
Such a description is appropriate for open quantum systems, for example, where the system being described is free to interact with other unmonitored degrees of freedom.
The unmonitored degrees of freedom appear to leech information out of the system being described and cause it to evolve nonunitarily.
In precise terms, a quantum channel is defined as follows.
\begin{defn}
Let $\Hil_A$ and $\Hil_B$ denote Hilbert spaces and let $\mathcal{L}(\Hil_A)$ and $\mathcal{L}(\Hil_B)$ denote the spaces of linear operators on $\Hil_A$ and $\Hil_B$, respectively.
A linear operator\footnote{A channel need not be linear, and we will indeed later encounter an example of a nonlinear channel. See \Ref{PreskillCh3} for further discussion of nonlinearity and some of its associated issues.} $\mathcal{N} : \mathcal{L}(\Hil_A) \rightarrow \mathcal{L}(\Hil_B)$ is a \emph{quantum channel} or \emph{superoperator} if it obeys the following conditions:
\begin{itemize}
\item[$i$.] $\mathcal{N}$ maps Hermitian operators onto Hermitian operators.
\item[$ii$.] $\mathcal{N}$ is trace-preserving.
\item[$iii$.] $\mathcal{N}$ is completely positive, i.e., for any extension of $\Hil_A$ to $\Hil_A \otimes \Hil_X$, the map $\mathcal{N} \otimes I_X$ is positive.
\end{itemize}
\end{defn}
\noindent Channels are conventionally defined as above so that they have an operator-sum representation, among other reasons.

In the case that $\Hil_A$ and $\Hil_B$ correspond to degrees of freedom held by two different parties, $A$ and $B$, a quantum channel can be thought of as a generalization of a classical communication channel that transmits quantum information from $A$ to $B$.
Just as one can ask what the capacity of a classical communication channel to transmit bits is, a natural question to ask is what the capacity of a quantum channel to transmit qubits is.
However, while Shannon's theorem \cite{Shannon1949} provides a clean expression for the capacity of a classical channel, there is no similarly universal and tidy expression for quantum channel capacity.

Intuitively, quantum channel capacity (which we hereafter refer to as just ``capacity'') is the ratio of the number of qubits transmitted by the channel to the number of qubits taken as input per use of the channel.
The capacity depends sensitively on the details of its definition. 
It depends as well as on what resources are available to the parties operating the channel, such as, for example, whether the parties $A$ and $B$ are allowed to communicate classically or share entangled ancillae that they can consume to assist their communication.

As an illustration, let us define the asymptotic channel capacity for parties that are unassisted by shared entanglement or classical communication.
(This definition is given in Sec.~10.7 of \Ref{Preskill:2016htv}.)
Let $\mathcal{N}^{A \rightarrow B}$ be a channel from $\Hil_A$ to $\Hil_B$, where we have introduced superscripts to indicate between which spaces the channel acts.
We introduce two additional Hilbert spaces, $\Hil_R$ and $\Hil_E$.
We define $\Hil_R$ to be a reference space, with dimension at most that of $\Hil_A$, such that any input to the channel, $\rho_A$, can be written as the reduced state of some pure state $\ket{\psi}_{RA} \in \Hil_R \otimes \Hil_A$.
In other words, for each $\rho_A$, there is a state $\ket{\psi}_{RA}$ such that
\begin{equation}
\rho_A = \Tr_R \ketbra{\psi}{\psi}_{RA}.
\end{equation}
Similarly, $\Hil_E$ is an environmental space onto which $\mathcal{N}^{A \rightarrow B}$ can be extended to an isometry $U^{A \rightarrow BE}$ such that, altogether, $I_R \otimes U^{A \rightarrow BE} \ket{\psi}_{RA} = \ket{\phi}_{RBE}$ maps a pure state onto another pure state.

We now make several auxiliary definitions. In terms of these additional Hilbert spaces, coherent information is defined as follows.
\begin{defn}
The \emph{coherent information} from $R$ to $B$ is
\begin{equation} \label{eq:Ic}
\begin{aligned}
I_c(R \rangle B)_\phi &\equiv -S(R|B) \\
&\equiv S(\rho_B) - S(\rho_{RB}) \\
&= S(\rho_{B}) - S(\rho_{E}).
\end{aligned}
\end{equation}
\end{defn}
\noindent The second line above is just the definition of conditional entropy, and the third line follows because $\ket{\phi}_{RBE}$ is a pure state.

Note that $I_c(R \rangle B)_\phi$ depends on neither the purification $\ket{\psi}_{RA}$ of $\rho_A$ nor the choice of dilation $U^{A \rightarrow BE}$ of $\mathcal{N}^{A \rightarrow B}$, as can be seen from the third and second lines of \Eq{eq:Ic}, respectively.
Coherent information is a measure of how much information makes it through the channel, in the sense that $I_c(R\rangle B)_\phi > 0$ means that the reference system $R$ is more correlated with $B$ than the environment $E$.
This is particularly evident if one rewrites $I_c(R \rangle B)_\phi$ in terms of mutual information:
\begin{equation}
I_c(R \rangle B)_\phi = \tfrac{1}{2}\left( I(R:B) - I(R:E) \right) .
\end{equation}
Coherent information therefore captures the amount of quantum information transmitted by a single use of the channel, which is formalized in the definition below.
\begin{defn}
The \emph{one-shot quantum channel capacity} is
\begin{equation}
Q_1(\mathcal{N}) \equiv \sup_A I_c(R \rangle B)_\phi,
\end{equation}
where the supremum is over all states $\rho_A$.
\end{defn}

Finally, we arrive at the definition of the asymptotic channel capacity by considering the limiting case in which the two parties are allowed multiple uses of the channel.
\begin{defn}
The \emph{quantum channel capacity} is 
\begin{equation}
Q(\mathcal{N}) \equiv \lim_{n \rightarrow \infty} \sup_{A^n} \frac{1}{n} I_c(R^n \rangle B^n)_{\phi_{R^n B^n E^n}}.
\end{equation}
\end{defn}
\noindent In analogy with the asymptotic definition of classical channel capacity, quantum channel capacity is therefore the average rate at which quantum information is transmitted over the channel, per channel use.

An important point is that asymptotic channel capacity is not in general equal to the one-shot capacity because channel capacity can be superadditive.
This is because, in many cases, quantum error correction and a cleverly designed communication protocol can allow the communicating parties to overcome some of the noisy losses incurred during use of the channel by redundantly encoding their messages over the course of several channel uses.
In other words, $n$ correlated uses of a channel can in general result in the transmission of more quantum information that $n$ uncorrelated, repeated uses of the channel.
The case of additive capacity, where $Q(\mathcal{N}) = Q_1(\mathcal{N})$, is a relatively special case.

An important  lesson to take from the formalism in this present section is that the notion of a ``quantum channel capacity'' requires a significant clarification in general before it is well defined. We must therefore carefully define the specific types of quantum channel capacities we want before we can apply the language of quantum channels to the traversable wormhole geometries we are interested in here.

\subsection{Channels from bipartite Hamiltonians}

We now focus on a particular class of quantum channels: those generated by bipartite unitary gates \cite{Bennett:2003}.
Consider a bipartite Hilbert space $\Hil = \Hil_A \otimes \Hil_B$, where $\dim \Hil_A = \dim \Hil_B = d < \infty$ and the factors $\Hil_A$ and $\Hil_B$ correspond to systems held by two parties, $A$ and $B$, respectively.\footnote{We can assume that the dimensions of the two factors are equal, without loss of generality, because the smaller of two Hilbert spaces can always be embedded into a larger space to match the dimension of the other factor.}
Let $U : \Hil \rightarrow \Hil$ be a unitary operator, which maps a joint state shared by $A$ and $B$ to another joint state.
However, given the bipartition of $\Hil$, one can think of $U$ as defining a two-way quantum channel between $A$ and $B$.
Via the action of $U$, information about the state held by $A$ propagates to $B$ and vice versa.
One very natural way such channels arise is through time evolution when the systems held by $A$ and $B$ are coupled.
In this case, $U$ is just the time evolution operator generated by the joint Hamiltonian on $\Hil_A \otimes \Hil_B$.

A basic quantity of interest for this setup is the channel capacity of the bipartite unitary operator $U$.
That is, how much quantum information can reliably be transmitted between the two parties via use of the channel?
While the precise calculation of asymptotic channel capacity is still a formidable task in this restricted setup, much is known about the entanglement capacity of such channels, which we elaborate on in the rest of this section.

\begin{figure}
\centering
\includegraphics[scale=1]{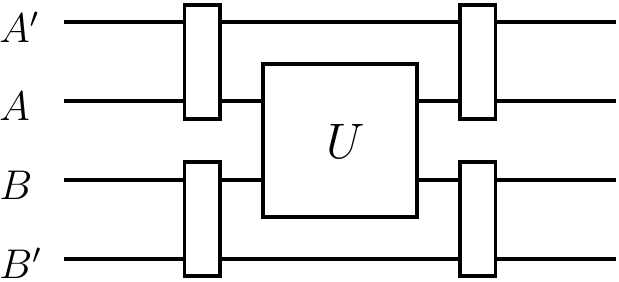}
\caption{The most general protocol that $A$ and $B$ can perform that uses $U$ exactly once and that makes use of their freely-available resources: local unitary operations, ancillae ($A^\prime$ and $B^\prime$), and classical communication. (The classical side channel is not explicitly shown.)}
\label{fig:1use}
\end{figure}

We will assume here that two-way classical communication is a free resource shared by $A$ and $B$.
Per \Ref{Bennett:2003}, we will also assume that $A$ and $B$ have access to ancillae of arbitrarily large (but finite) dimension and that $A$ and $B$ are allowed to perform local unitary operations.
Consequently, denoting the ancillary systems by $A^\prime$ and $B^\prime$, the most general protocol that uses $U$ once is shown in \Fig{fig:1use} (cf. \Ref{Bennett:2003}). 

\subsection{Entanglement capacity}

Loosely speaking, entanglement capacity quantifies the ability of a channel to generate entanglement.
We will be interested in entanglement production between two parties, so here we will consider the case in which the channel maps between states in the same Hilbert space, where the Hilbert space decomposes into two factors corresponding to the two parties that become entangled.

We will follow the definitions and notation of \Ref{Bennett:2003}.
Let $\N : \mathcal{L}(\Hil) \rightarrow \mathcal{L}(\Hil)$ be a channel between states on the Hilbert space $\Hil$, and suppose that $\Hil$ decomposes into the tensor product $\Hil = \Hil_A \otimes \Hil_B$ with $\dim \Hil_A = \dim \Hil_B = d < \infty$.
Then an entanglement capacity can be defined as follows.

\begin{defn}\label{def:tshot}
The \emph{$t$-shot entanglement capacity} of $\N$ with respect to the entanglement measures $E_{\mrm{in}}$ and $E_\mrm{out}$ is
\begin{equation}
E_{\mrm{in}\rightarrow\mrm{out}, \, \mathcal{N}}^{(t,\varnothing,r)} \equiv \sup_{\mathcal{P}_t} \frac{1}{t} \left[ E_\mrm{out}\left(\mathcal{P}_t(\ketbra{00}{00}) \right) - E_\mrm{in}\left(\ketbra{00}{00}\right) \right].
\end{equation}
The supremum is over all protocols $\mathcal{P}_t$ that use $\mathcal{N}$ $t$ times.
The argument $r$ denotes the collection of freely available resources, such as local unitary operations and classical communication (LOCC), as well as ancillae.
The empty set symbol, $\varnothing$, specifies that the initial state is chosen to be an unentangled state between $\Hil_A$ and $\Hil_B$, denoted by $\ket{00}$.
(Here we omit subscripts on the state.)
In the case where we are free to prepare any initial state, the capacity is given by
\begin{equation}
E_{\mrm{in}\rightarrow\mrm{out}, \, \mathcal{N}}^{(t,*,r)} \equiv \sup_{\mathcal{P}_t} \sup_{\rho} \frac{1}{t} \left[ E_\mrm{out}\left(\mathcal{P}_t(\rho) \right) - E_\mrm{in}\left(\rho\right) \right].
\end{equation}
\end{defn}
\noindent Note that we will always assume that LOCC is a freely available resource, in which case the choice of initial unentangled state does not matter. Note also that the asymptotic limit is denoted
\begin{equation}
E_{\mrm{in}\rightarrow\mrm{out}, \, \mathcal{N}}^{(*,r)} = \lim_{t \rightarrow \infty} E_{\mrm{in}\rightarrow\mrm{out}, \, \mathcal{N}}^{(t,*,r)} \, .
\end{equation}

Definition \ref{def:tshot} makes clear that entanglement capacity depends on the choice of entanglement measures used to quantify the entanglement between $\Hil_A$ and $\Hil_B$.
Abstractly, an entanglement measure is defined as follows \cite{Bennett:2003}.

\begin{defn}
An \emph{entanglement measure} on states in $\Hil = \Hil_A \otimes \Hil_B$ is a function $E : \mathcal{L}(\Hil) \rightarrow [0,\infty)$ that obeys the following conditions:
\begin{itemize}
\item[$i$.] $E(\ketbra{\Psi}{\Psi}) = 0$ for product states $\ket{\Psi} = \ket{\psi}_A \otimes \ket{\phi}_B$.
\item[$ii$.] $E$ is invariant under local unitaries, i.e.,
\begin{equation*}
E(U_A \otimes U_B \, \rho \, U_A^\dagger \otimes U_B^\dagger) = E(\rho).
\end{equation*}
\item[$iii$.] $E$ is nonincreasing under LOCC.
\item[$iv$.] For all states $\rho$, $E(\rho \otimes \ketbra{\Phi_d}{\Phi_d}) = E(\rho) + E(\ketbra{\Phi_d}{\Phi_d})$, where $\ket{\Phi_d}$ denotes the maximally entangled state across $A$ and $B$,
\begin{equation*}
\ket{\Phi_d} = \frac{1}{\sqrt{d}} \sum_{i=1}^d \ket{i}_A \ket{i}_B \, .
\end{equation*}
\end{itemize} 
\end{defn}

\noindent For example, the entanglement entropy with respect to one of the factors, say $A$,
\begin{equation}
E_e(\rho) = - \Tr \rho_A \log_2 \rho_A \, ,
\end{equation}
is an entanglement measure, where $\rho_A = \Tr_B \rho$ denotes the reduced state of $\rho$ on $A$.

Two other entanglement measures that we will consider here are the entanglement cost, $E_c$, and the distillable entanglement, $E_d$.
The entanglement cost of a state $\rho$ is essentially the number of Bell pairs that $A$ and $B$ must consume in order to prepare the state $\rho$ using only LOCC.
Its precise definition is as follows \cite{Hayden:2001}.

\begin{defn}
The \emph{entanglement cost} of a state $\rho$ is defined as
\begin{equation}
\begin{aligned}
E_c(\rho) \equiv \inf \left\{ E ~ \vert ~ \right. & \left. \forall \, \epsilon > 0, \delta > 0, ~ \exists \, m, n, \mathcal{N} \right. \\
& \left. \mrm{such}~\mrm{that} ~ \left| E - \tfrac{m}{n} \right| \leq \delta ~ \mrm{and} ~ D\left(\mathcal{N}(\ketbra{\Psi^-}{\Psi^-}^{\otimes m}),\rho^{\otimes n}\right) \leq \epsilon  \right\} \, .
\end{aligned}
\end{equation}
In the above, $\ket{\Psi^-}$ is a single copy of a Bell singlet state shared between $A$ and $B$,
\begin{equation}
\ket{\Psi^-} = \frac{1}{\sqrt{2}} \left( \ket{01} - \ket{10}  \right),
\end{equation}
$\mathcal{N}$ is any LOCC channel acting on $m$ copies of $\ket{\Psi^-}$, and $D$ is the Bures distance,
\begin{equation}
D(\rho,\rho^\prime) = 2 \sqrt{1-F(\rho,\rho^\prime)},
\end{equation}
where $F(\rho,\rho^\prime) = \Tr \sqrt{\rho^{1/2} \rho^\prime \rho^{1/2}}$ is the Uhlmann fidelity.
\end{defn}
In other words, $E_c(\rho)$ quantifies the asymptotic rate at which Bell pairs are consumed to produce copies of $\rho$.
That is, if $A$ and $B$ must use $m$ shared Bell pairs to produce $n$ copies of $\rho$ as $m$ and $n$ grow large, then $E_c$ expresses the fact that each copy of $\rho$ ``costs'' $m/n$ shared Bell pairs.

Similarly, the distillable entanglement $E_d(\rho)$ of a state $\rho$ is essentially the number of Bell pairs that $A$ and $B$ can extract from the state $\rho$ using only LOCC.
Its precise definition is similar to the definition of $E_c$ above.

\begin{defn}
The \emph{distillable entanglement} of a state $\rho$ is defined as
\begin{equation}
\begin{aligned}
E_d(\rho) \equiv \sup \left\{ E ~ \vert ~ \right. & \left. \forall \, \epsilon > 0, \delta > 0, ~ \exists \, m, n, \mathcal{N} \right. \\
& \left. \mrm{such}~\mrm{that} ~ \left| E - \tfrac{m}{n} \right| \leq \delta ~ \mrm{and} ~ D\left(\ketbra{\Psi^-}{\Psi^-}^{\otimes m},\mathcal{N}(\rho^{\otimes n})\right) \leq \epsilon  \right\} \, .
\end{aligned}
\end{equation}
\end{defn}
In other words, $E_d$ quantifies the asymptotic rate at which Bell pairs can be distilled if $A$ and $B$ share many copies of a given state $\rho$.
Note that $E_c$, $E_d$, and $E_e$ all coincide when $\rho$ is a pure state \cite{2005quant.ph..4163P}.

Armed with these definitions, a natural measure of the ability of a channel to generate entanglement is therefore the entanglement capacity with $E_\mrm{in} = E_c$ and $E_\mrm{out} = E_d$; this capacity measures the ability of a channel to yield a net gain (or loss) of Bell pairs.

The importance of the entanglement capacity $E_{c \rightarrow d, \, \mathcal{N}}^{(*,r)}$ (or $E_{c \rightarrow d, \, \mathcal{N}}^{(\varnothing,r)}$) is that it provides a lower bound for the channel capacity via the following explicit protocol \cite{Bennett:1996gf}.
Asymptotically, each use of the channel produces at most $E_{c \rightarrow d, \, \mathcal{N}}^{(*,r)}$ clean Bell pairs; given the channel output, one can perform an entanglement purification protocol to extract at most $E_d$ shared Bell pairs between $A$ and $B$, but $E_c$ Bell pairs must be consumed to generate the input for the next run of the channel.
Since $A$ and $B$ share a classical communication channel, they can use the newly produced $E_{c \rightarrow d, \, \mathcal{N}}^{(*,r)}$ Bell pairs to run a teleportation protocol \cite{PhysRevLett.70.1895}.
Recall that teleportation consumes $E_{c \rightarrow d, \, \mathcal{N}}^{(*,r)}$ Bell pairs and $2 E_{c \rightarrow d, \, \mathcal{N}}^{(*,r)}$ bits of classical communication to transfer an arbitrary state of $E_{c \rightarrow d, \, \mathcal{N}}^{(*,r)}$ qubits from $A$ to $B$.
Since we have exhibited an explicit protocol which, through use of the channel $\mathcal{N}$, achieves an asymptotic qubit transfer rate of $E_{c \rightarrow d, \, \mathcal{N}}^{(*,r)}$, then it follows that the channel capacity of $\mathcal{N}$ must be at least as big as $E_{c \rightarrow d, \, \mathcal{N}}^{(*,r)}$, i.e.,
\begin{equation}
Q(\mathcal{N}) \geq E_{c \rightarrow d, \, \mathcal{N}}^{(*,r)} \, .\label{eq:QgtrE}
\end{equation}
 
\subsection{Entanglement capacity of bipartite unitary channels}

A key result of \Ref{Bennett:2003} is that many entanglement capacities for bipartite unitary channels are additive and independent of classical communication, which we denote by ``cc''.
In particular, it is shown therein that
\begin{equation}
E_{c \rightarrow d, U}^{(t,*,\text{cc})} = E_{c \rightarrow d, U}^{(1,*)} \, .\label{eq:oneshotequalsmultishot}
\end{equation}
Therefore, an optimal protocol for generating entanglement that uses $U$ $t$ times is just to use an optimal one-use protocol $t$ times.
Moreover, via a host of corollaries, it is further shown that
\begin{align}
\label{eq:cor1} E_{c \rightarrow d, U}^{(1,*)} &= E_{c \rightarrow c, U}^{(1,*)} \\[2mm]
\label{eq:cor2} E_{c \rightarrow c, U}^{(1,*)} &= \sup_{\ket{\psi}} E_c(U\ket{\psi}) - E_c(\ket{\psi}) \\[2mm]
\label{eq:cor3} E_{c \rightarrow c, U}^{(1,*)} &= E_{c \rightarrow c, U}^{(\varnothing)}.
\end{align}
The corollaries \eqref{eq:cor1} and \eqref{eq:cor2} imply that the optimal one-shot protocol can be realized with a pure input state, and corollary \eqref{eq:cor3} establishes that the asymptotic entanglement capacities with and without the ability to prepare arbitrary input states (the resource $*$) are equal.
Note, however, that the one-shot capacity $E_{c\rightarrow d,U}^{(1,\varnothing)}$ may be different.

\section{Review of Traversable Wormholes in AdS/CFT}\label{sec:AdSCFT}

Having carefully defined our quantum information-theoretic quantities of interest, we now turn to the specific system under consideration: the holographic traversable wormhole. In this section, we briefly review the geometrical arguments of Refs.~\cite{Gao:2016bin,Maldacena:2017axo}, which show that a double-trace deformation of the thermofield double state for the boundary CFTs leads to traversability of the wormhole in the bulk holographic description. 
We will furthermore use the machinery of Refs.~\cite{Dray:1984ha,Shenker:2013pqa} to explicitly connect the size of the deformation's coupling to the amount of negative energy falling towards the wormhole and hence the amount by which the horizon is shifted and the wormhole rendered traversable.

The thermofield double state in the tensor product of two identical noninteracting theories is the state that results in a thermal density matrix at inverse temperature $\beta$ if either of the two theories is traced out:
\be
\ket{\Psi} = \frac{1}{\sqrt{Z}} \sum_n e^{-\beta E_n/2} \ket{\bar n}_L \otimes \ket{n}_R,
\ee
where $Z$ is the temperature-$\beta^{-1}$ partition function of one of the non-interacting theories \cite{Maldacena:2001kr}, a bar denotes CPT conjugation, and $\ket{n}$ and $E_n$ denote the energy eigenstates and eigenvalues, respectively, of each theory.
If we specialize to the case in which both theories are large-$N$ CFTs on the $(D-1)$-dimensional boundary sphere  (or, equivalently, the case in which each theory lives on a separate boundary sphere), the bulk description of $\ket{\Psi}$ is the two-sided AdS-Schwarzchild black hole, with metric
\be
{\rm d}s^2 = -f(r){\rm d}t^2 + \frac{{\rm d}r^2}{f(r)}+r^2 {\rm d}\Omega^2_{D-2},
\ee
where
\be 
f(r)\equiv 1 - \frac{16 \pi G_D M_D}{(D-2)\Omega_{D-2}r^{D-3}} -\frac{2\Lambda}{(D-1)(D-2)}r^2.\label{eq:deff}
\ee
Here, $\Lambda$ is the cosmological constant, ${\rm d}\Omega^2_{D-2}$ is the angular metric, $\Omega_{D-2}$ is the volume of the $(D-2)$-sphere, and $M_D$ is a mass parameter corresponding to the mass of a black hole with temperature $\beta$. It is often convenient to define the AdS length as $\ell \equiv \sqrt{-(D-1)(D-2)/2\Lambda}$, so that the last term in \Eq{eq:deff} becomes simply $+r^2/\ell^2$.

\begin{figure}
\centering
\includegraphics[width = 13cm]{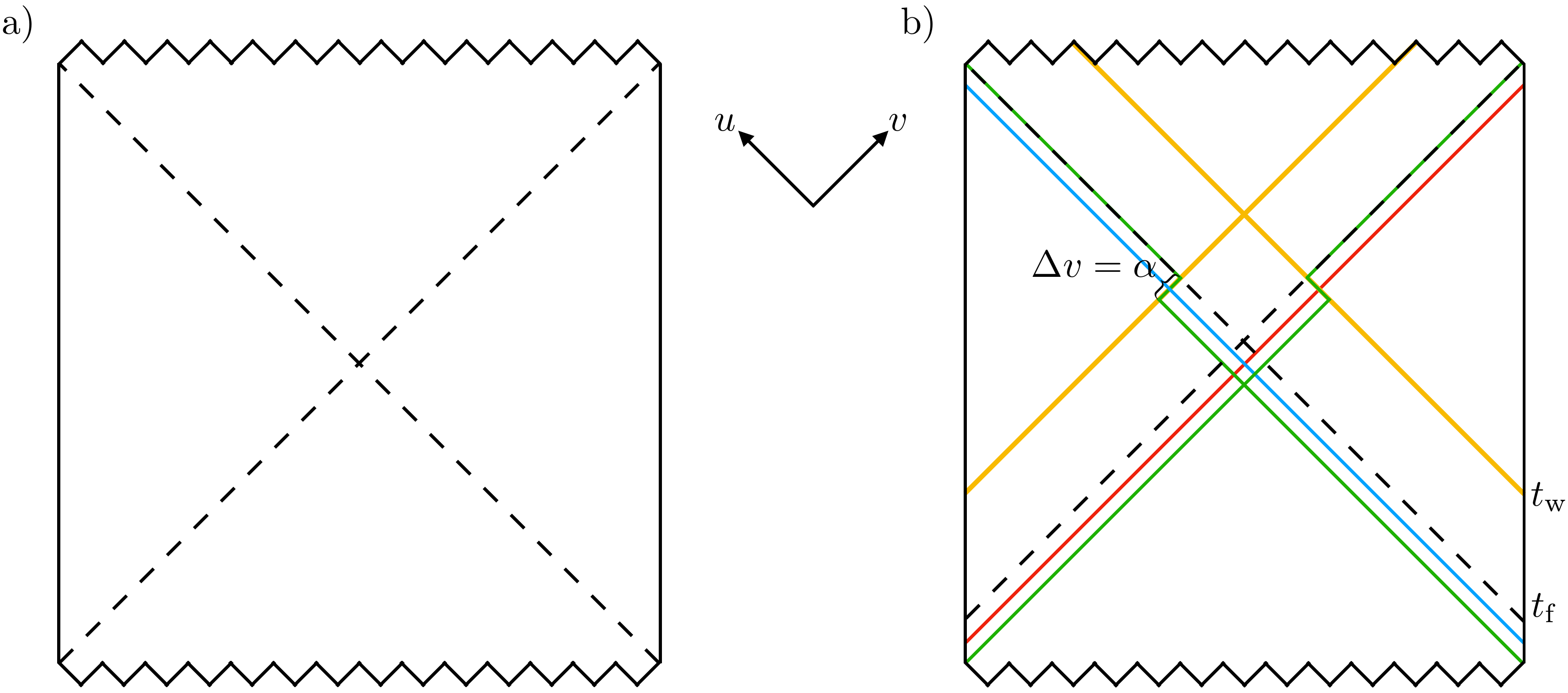}
\caption{Penrose diagrams for the AdS wormhole, with future event horizons illustrated by the black dashed lines. In panel a), the standard AdS black hole geometry is depicted, which in $D=3$ dimensions has the BTZ metric given in \Eqs{eq:BTZ}{eq:BTZuv}. In panel b), the spacetime has been modified by the double-trace deformation at time $t_{\rm w}$, resulting in a negative null energy shock wave in the bulk (yellow line). This shock wave has the effect of moving the apparent horizon inward (green line), rendering the wormhole traversable: signals sent from the left boundary before $t_{\rm f}$ can reach the right boundary (red line) and vice versa (blue line).}
\label{fig:geometry}
\end{figure}

Following Refs.~\cite{Shenker:2013pqa,Gao:2016bin}, let us specialize to the nonrotating BTZ black hole in $D=3$ dimensions \cite{Banados:1992wn}. Defining a unitless mass $m\equiv8G_3 M\equiv8G_3 M_3-1$ proportional to the ADM mass $M$ of the geometry \cite{Carlip:1995qv}, \Eq{eq:deff} becomes simply $-m+(r^2/\ell^2)$, so we can write the metric as
\be
{\rm d}s^2 = -\frac{r^2-r_{\rm h}^2}{\ell^2}{\rm d}t^2 + \frac{\ell^2}{r^2-r_{\rm h}^2}{\rm d}r^2 + r^2 {\rm d}\phi^2,\label{eq:BTZ}
\ee
where $r_{\rm h} = \ell \sqrt{m}$. To avoid a naked conical singularity, $m$ must be nonnegative so that $r_{\rm h}$ is real (except for the case of $m=-1$, which corresponds to pure AdS \cite{Banados:1992wn}). In Kruskal coordinates $(u,v)$ defined by $e^{2r_{\rm h} t/\ell^2} = -v/u$ and $r/r_{\rm h} = (1-uv)/(1+uv)$ in the right wedge, this becomes
\be
{\rm d}s^2=\frac{-4\ell^2 {\rm d}u {\rm d}v + r_{\rm h}^2 (1- u v)^2 {\rm d}\phi^2}{(1+uv)^2}.\label{eq:BTZuv}
\ee
In these coordinates, which can be analytically continued to cover the entire two-sided geometry, the past and future singularities are located at $uv=+1$, the horizons are located at $uv=0$, and the two boundaries are located at $uv=-1$. The geometry is sketched in part a) of \Fig{fig:geometry}.

It is clear that the metric (\ref{eq:BTZ}) describes a (marginally traversable) wormhole geometry, in which particles falling from one exterior across the horizon are unable to escape into the other exterior.
From the bulk perspective, the wormhole could be rendered traversable by sending in a null energy condition-violating shock wave.
The question is whether such a shock wave can be naturally created by operators in the boundary theory.
In \Ref{Gao:2016bin}, a natural-seeming double-trace deformation of the boundary was considered, in which relevant operators dual to bulk scalars are entangled across the two theories, giving an effectively bilocal contribution to the action:
\be
\delta S = \int {\rm d}t \,{\rm d}\phi\, h(t,\phi)\mathcal{O}_R(t,\phi)\mathcal{O}_L(-t,\phi),\label{eq:dbltraceaction}
\ee 
where $\mathcal O$ has conformal dimension $\Delta$ and the coupling $h$ has support only in some time window.  The minus sign appearing as an argument in ${\cal O}_L$ is present because $t$ is the time associated with the bulk timelike Killing vector, which runs in opposite directions in the left and right wedges; hence, $\delta S$ as constructed turns on the double-trace deformation at the same boundary time as seen in the CFT.

For positive $h$, the integrated energy falling through the horizon $\int {\rm d}u\, T_{uu}|_{v=0}$ is negative: in detail, $T_{uu}$ is initially negative once the pulse has had time to reach the horizon, and though it later becomes positive, the integrated energy flux remains negative \cite{Gao:2016bin}.
The resulting geometry is shown in part b) of \Fig{fig:geometry}; probes sent from the boundary towards the origin of the spacetime at times earlier than the deformation is applied have a window in which they can escape into the other exterior region and eventually be received on the other boundary.
In later sections we will exploit this ability to send some signals from one boundary to the other for information-theoretic purposes.
In the remainder of this section, we confine ourselves to working out (an approximation to) the deformed metric, specifically the width $\Delta v$ of the window in which signals can escape into the second asymptotic region.

In principle, we could work out the post-deformation metric\footnote{Of course, a generic CFT perturbation around the TFD state might not necessarily have a classical bulk description. We are assuming here that the double-trace deformation is such that in any description of a single asymptotic region only classical matter is added to the boundary, even though the deformation must in general change the entanglement structure between the two regions since it changes the structure between the two boundaries. This picture might not be fully self-consistent, in which case we could, for example, resort to the semiclassical description and compute the metric sourced by the expectation value of the stress-energy tensor operator. See \Ref{jafferis} for related discussion of the validity of a bulk geometric description.} by first evaluating the Green function for $\phi$ in the modified boundary conditions sourced by the deformation \cite{Witten:2001ua,Hartman:2006dy}, then computing the stress tensor, and finally inverting the linearized (or full) Einstein equations to find the new metric. 

For the restricted shock wave-type problem relevant to us, however, we can take a simpler approach. 
The methods of Refs.~\cite{Dray:1984ha,Shenker:2013pqa} apply to problems in which a solution to the vacuum Einstein equations is deformed by a delta-function perturbation at the origin.
Provided some consistency conditions are satisfied, the leading change to the metric is a shift in the location of the horizon.
In particular, \Ref{Shenker:2013pqa} considers the case of shock waves in AdS, i.e., an AdS-Vaidya geometry.
Suppose we release a shock wave of boundary energy $E$ at Killing time $t_{\rm w}$, thereby taking a BTZ solution with mass $M$ in the far past to one with mass $M+E$ in the far future. For the double-trace coupling $h$, we have
\be
E \sim - h G_3 M/\ell. \label{eq:hErelation}
\ee
The AdS-Vaidya geometry that glues these two spacetimes together along the null surface of the shock wave is \cite{Shenker:2013pqa}
\be
{\rm d}s^2 = \frac{1}{(1+u \hat{v})^2} \[ -4 \ell^2 {\rm d}u {\rm d}\hat{v} + 4 \ell^2 \alpha \delta(u) {\rm d}u^2 + r_{\rm h}^2 (1-u\hat{v})^2 {\rm d}\phi^2\],
\ee
where the hatted coordinate is given by $\hat v = v - \alpha \theta(u)$ and
\be
\alpha \equiv - \frac{E}{4M} e^{r_{\rm h} t_{\rm w}/\ell^2} = {\cal O}(1)\times \frac{h G_3}{\ell} e^{r_{\rm h} t_{\rm w}/\ell^2},\label{eq:alpha}
\ee
where the ${\cal O}(1)$ factor depends on the time-dependent profile one uses for the double-trace coupling $h$ in \Eq{eq:dbltraceaction}. This expression is exact for fixed $\alpha$ in the limit where $E/M\rightarrow 0$ and $t_{\rm w} \rightarrow \infty$ simultaneously \cite{Shenker:2013pqa}. In our case, $E$ is negative, so the null energy condition is violated, allowing the wormhole to be traversable; in particular, $T_{uu} = -(\alpha/4\pi G_3)\delta(u)$.

In our case, where we are considering the double-trace deformation, we by construction have two shocks, one approaching from the right and one from the left.
Hence we should similarly replace $u$ with $\hat u = u - \alpha \theta(v)$ and the metric is
\be
{\rm d}s^2 = \frac{1}{(1+\hat{u} \hat{v})^2} \[ -4 \ell^2 {\rm d}\hat{u} {\rm d}\hat{v} + 4 \ell^2 \alpha \(\delta(\hat{u}) {\rm d}\hat{u}^2 + \delta(\hat{v}) {\rm d}\hat{v}^2\) + r_{\rm h}^2 (1-\hat{u}\hat{v})^2 {\rm d}\phi^2\].\label{eq:superimposed_shockwaves}
\ee
That is, both horizons are shifted inward in the Kruskal coordinates by $\alpha$ for $E$ negative. In the regime we are considering, in which gravitational interactions between the two shock waves can be neglected, the two shocks can be simply superimposed, as in \Eq{eq:superimposed_shockwaves}.

We have arrived at an expression (\ref{eq:alpha}) for the horizon shift in terms of the energy of the shock waves created at the boundary by the double-trace deformation. We will now discuss in what ways wormhole traversability can and cannot be interpreted in information-theoretic terms, in particular in the language of entanglement witnesses and quantum channels.

\section{Traversable Wormholes Do Not Make Entanglement an Observable}\label{sec:observable}

It might appear that the procedure described in the previous section for rendering wormholes traversable makes it possible to determine whether any pair of black holes is connected by a wormhole: one could simply assume that such a wormhole exists, perform the appropriate double-trace deformation to make the wormhole traversable, and send a signal into one member of the pair and check whether it emerges from the other. 
More explicitly, within the AdS/CFT setup described in \Sec{sec:AdSCFT}, we could imagine that the experimenter has access to a number of CFT boundaries dual to bulk black hole geometries and wishes to check if a particular pair of black holes is connected by a wormhole or, equivalently, if two boundaries are connected by a quantum channel like the ones we will describe in \Sec{sec:TWQC}.

If this experiment could be performed with perfect reliability, so that it was always possible to verify that two black holes were connected by a wormhole, it would violate a fundamental principle of quantum mechanics, namely, that entanglement is not an observable. 
More precisely, because a superposition of states that are entangled (in some basis) need not itself be entangled, the set of all entangled states in a bipartite Hilbert space $\mathcal H = \mathcal A \otimes \mathcal B$, $\mathcal E \equiv \{\ket{\Psi}:S(\Tr_{\mathcal A} \ket{\Psi}\bra{\Psi})\ne 0\}$, is not a subspace of the Hilbert space. 
Hence, linearity of quantum mechanics requires that no projector onto $\mathcal E$ exists, so entanglement is not a quantum-mechanical observable.
More generally, no subset of $\mathcal E$ (except for trivial subsets consisting of single entangled states) is itself a subspace and so, while a projector onto any individual entangled state exists, there is no such projector onto a set of more than one entangled state.

As was pointed out in \Ref{Bao:2015nca},\footnote{For the most part, Ref. \cite{Bao:2015nca} worked within the context of the ER/EPR hypothesis, in which {\it every} entangled state is meant to be connected by a (perhaps microscopic or highly quantum) wormhole. 
We have restated the argument of \Ref{Bao:2015nca} in a form that does not rely on the ER/EPR conjecture.} because wormhole geometries are described by entangled states such as the thermofield double state, there is no quantum-mechanical observable that can differentiate between such states and the entire collection of product states of the two boundaries (which includes, e.g., states that describe an unentangled black hole in each bulk region).
This is the holographic consequence of entanglement not being an observable.
Of course, given a particular entangled state, such as the thermofield double, one can distinguish it from a particular product state by measuring some operator.
However, a particular entangled state cannot be distinguished from an unknown product state, and much less an unknown entangled state from an unknown product state.
In other words, given a particular entangled state and an operator, there is always a separable state that reproduces the entangled state's measurement statistics of that operator.

Hence there must be a gravitational obstruction in the bulk that prevents any procedure from determining with perfect reliability whether a bulk geometry containing a black hole is connected by a wormhole to a different bulk region.
\Ref{Bao:2015nca} treated standard nontraversable AdS wormholes and hence considered procedures in which a bulk observer crossed the horizon and looked for a signal (or another observer) in the black hole interior originating from a different asymptotic bulk geometry. It was argued that there always exist wormhole geometries in which the bulk observer would hit the black hole singularity before being able to receive any signals, so no completely reliable procedure for detecting the presence of a wormhole could exist. That is, since the metric exterior to the event horizon is time-independent, it is possible for the black hole to indeed be connected to another by a nontraversable wormhole, but  for this fact to be undetectable if the observer jumps into the black hole too late to observe a particular, fixed signal entering the wormhole from the other side (see Fig.~4 of \Ref{Bao:2015nca}).

\begin{figure}
\centering
\includegraphics[width = 13cm]{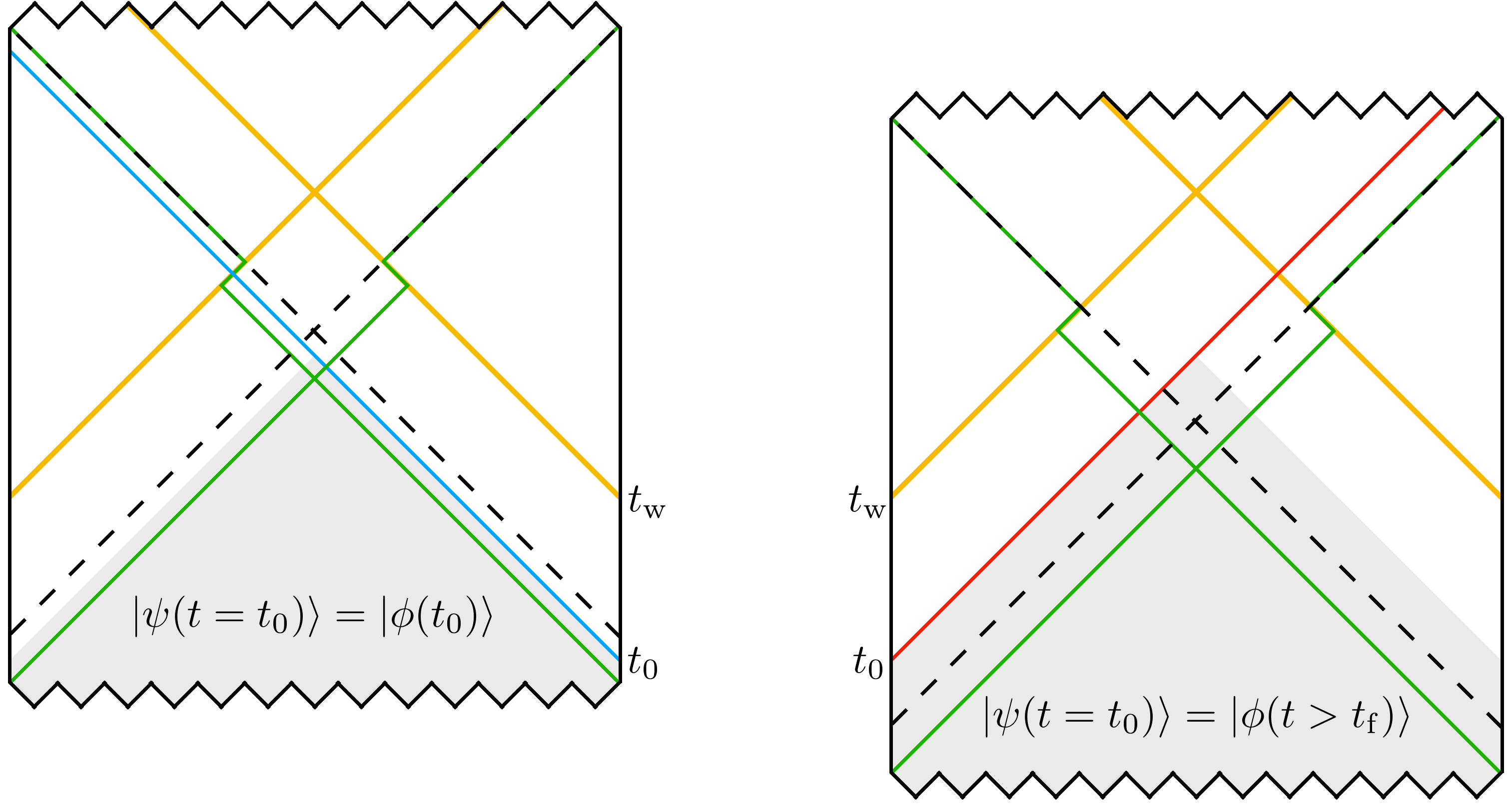}
\caption{Different outcomes for the double-trace deformation experiment, depending on the state $\ket{\psi(t_0)}$ at time $t_0$, which is dual to the geometry in the corresponding Wheeler-deWitt patch (shaded gray). The times of the signal $t_0$ (blue and red lines) and the double-trace deformation (yellow lines) are fixed, but the time coordinate of the geometry can be shifted up and down, due to the time-invariance of the AdS black hole metric. If $\ket{\psi(t_0)}$ corresponds to the thermofield double state at time $t_0$, $\ket{\phi(t_0)}$ (left), or more generally to any $\ket{\phi(t<t_{\rm f})}$, then the signal succeeds in traversing the wormhole (blue line, left). On the other hand, if the state $\ket{\psi(t_0)}$ corresponds to any $\ket{\phi(t>t_{\rm f})}$, the signal fails to traverse the wormhole, hitting the singularity (red line, right). The existence of states where such failures occur is a necessary consequence of the linearity of quantum mechanics, since entanglement is not an observable.}
\label{fig:shift}
\end{figure}

Making the wormholes traversable using the procedure of Refs.~\cite{Gao:2016bin, Maldacena:2017axo} allows a broader class of experiments in the bulk, in particular, the experiments mentioned above and depicted in panel b) of \Fig{fig:geometry}, in which no event horizon is actually crossed.
However, it does not change the quantum-mechanical argument forbidding a projector onto collections of entangled states, so there must be some bulk geometric circumstances under which the procedure can fail to send a signal through the wormhole.
In particular, recall \cite{Maldacena:2013xja} that the CFT states dual to a wormhole in the bulk are not restricted to only the thermofield double, but include a one-parameter family of states indexed by boundary time,
\be
\ket{\phi(t)} = \frac{1}{\sqrt{Z}} \sum_n e^{-\beta E_n/2} e^{-2iE_n t} \ket{\bar n}_L \otimes \ket{n}_R.\label{eq:states_t_shift}
\ee
These describe states of two entangled black holes that have both evolved forward by a time $t$ relative to the thermofield double state. 
As $t$ increases, the causal diamond extending into the bulk from the CFT boundary moves toward the future, with its intersection with the past singularity first decreasing and then eventually its intersection with the future singularity increasing. As discussed in \Ref{Bao:2015nca}, observers starting any finite distance away from the horizon in a wormhole geometry corresponding to a sufficiently large value of $t$ would hit the singularity before crossing into the other half of the spacetime.

A similar phenomenon, depicted in \Fig{fig:shift}, occurs here. Consider performing the double-trace deformation at fixed boundary time $t_{\rm w}$, on a wormhole geometry where the observer releases a signal from the boundary at some fixed time $t_0$. The state of the boundary at time $t_0$, $\ket{\psi(t_0)}$, corresponds to the Wheeler-deWitt patch in the bulk anchored to time $t_0$ on the boundary. Outside of the apparent horizon and to the past of the shock wave associated with the double-trace deformation, the geometry is still simply described by the time-independent exterior of the AdS black hole. Hence, $\ket{\psi(t_0)}$ can be any one of the $\ket{\phi(t)}$ in \Eq{eq:states_t_shift}, all of which are indistinguishable to boundary observers at $t_0$. If $\ket{\psi(t_0)} = \ket{\phi(t< t_{\rm f})}$, the signal makes it through the wormhole, as shown in the left panel of \Fig{fig:shift}, but if $\ket{\psi(t_0)} = \ket{\phi(t> t_{\rm f})}$, as in the right panel of \Fig{fig:shift}, the signal fails to traverse the wormhole and instead hits the singularity.
Thus, there is always some class of wormhole states where the signaling procedure would fail, and these states cannot be distinguished from two unentangled black holes.

Just as quantum mechanics does not forbid an observable that determines whether an arbitrary state is a member of a particular proper \emph{subset} of all entangled states, a successful traversal of the wormhole by a particular signal allows us to conclude that two black holes are entangled in a particular manner.\footnote{In particular, the geometrical information alone would allow us to single out the value of $t$ among the states in \Eq{eq:states_t_shift} corresponding to $\ket{\psi(t_0)}$, e.g., by measuring the curvature at some identifiable point in the spacetime (such as where the signal intersects the shock wave, cf. \Ref{Bao:2015nca}) or the total time delay between $t_0$ and the signal's reception on the other boundary (which depends on the redshift factor of the metric over the entire path of the signal, which in turn will be different for various $\ket{\phi(t)}$).} We will make this notion precise by using the tools of entanglement witnesses in Secs.\ \ref{sec:witness} and \ref{sec:TWEW} below, but we first consider more carefully the quantum-mechanical process that corresponds to rendering the wormhole traversable.

\section{Traversable Wormholes as Quantum Channels}\label{sec:TWQC}

Let us begin by revisiting the process of making a wormhole traversable and sending a bulk excitation through it, but from the perspective of the boundary theory.
Let $\Hil = \Hil_L \otimes \Hil_R$ denote the joint Hilbert space of two CFTs, which we refer to as the ``left'' and ``right'' CFTs, and suppose that we prepare the thermofield double state at some initial (boundary) time $t_{\rm i}$.
The basic procedure begins with acting at the left spacetime boundary with an operator $\phi_L$, which, from the perspective of the bulk, causes an excitation to begin propagating in toward the black hole.
Then, at a later time $t_\mrm{w}$, the double-trace deformation $\oh_L \oh_R$ is performed across both CFTs, which produces the negative null energy shocks in the bulk that make the wormhole traversable.
The end result is that the excitation produced by $\phi_L$ manifests itself in the right CFT at some later time $t_{\rm f}$.
From the perspective of the bulk, this is the time at which the excitation, having traversed the wormhole, reaches the right boundary.

From the perspective of the boundary theory, the entire process above is described by the unitary evolution of a state at $t_{\rm i}$ to a state at $t_{\rm f}$,
\begin{equation} \label{eq:fullUnitary}
\ket{\Psi(t_{\rm f})} = \mathcal{U}(t_{\rm f},t_{\rm w}) e^{ih\oh_L \oh_R} \mathcal{U}(t_{\rm w},t_{\rm i}) \phi_L \ket{\TFD(t_{\rm i})} .
\end{equation} 
The operator $\mathcal{U}(t_1,t_2)$ denotes the unitary time evolution operator derived from the CFT Hamiltonian that evolves a state in $\Hil$ from the time $t_1$ to $t_2$.

\subsection{A channel between boundary (sub)regions}\label{sub:channel_subregions}

The relation in \Eq{eq:fullUnitary} naturally gives rise to a quantum channel between the two CFTs.
We can think of the total time evolution from $t_{\rm i}$ to $t_{\rm f}$ as a bipartite unitary map, which generates a channel between the two CFT sides, as discussed in \Sec{sec:channels}.
Instead of just considering a map from $\Hil_L$ to $\Hil_R$, however, we can more generally consider maps from subfactors of $\Hil_L$ to subfactors of $\Hil_R$ that correspond to boundary subregions.
We do so on physical grounds: if our aim is to study how excitations created by $\phi_L$ propagate through the bulk and these excitations are created near the boundary, then from the perspective of the CFT it makes sense to think of these excitations as (initially) being localized to the minimal boundary subregions that contain them.
Of course, we can always take the boundary subregions to be the entire left and right CFTs to restore a channel between the full boundaries.

Given a boundary subregion $A$ in the left CFT and a subregion $B$ in the right CFT, the channel maps an initial state on $A$, obtained by acting with $\phi_L$ on $\ket{\TFD(t_{\rm i})}$ and tracing out $\bar A$, to the final reduced state on $B$ at $t_{\rm f}$.
In other words, we can characterize the channel, $\mathcal{N}^{A \rightarrow B}$, as follows:
\begin{equation} \label{eq:infchannel}
\begin{array}{rcl}
\mathcal{N}^{A \rightarrow B} \; : \; D(\mathcal{N}^{A \rightarrow B}) & \rightarrow  & \mathcal{L}(\Hil_B) \\[2mm]
\rho_A & \mapsto & \Tr_{\bar{B}} \left( \ketbra{\Psi(t_{\rm f})}{\Psi(t_{\rm f})} \right) .
\end{array}
\end{equation}
The domain of $\mathcal{N}^{A \rightarrow B}$, $D(\mathcal{N}^{A \rightarrow B}) \subset \mathcal{L}(\Hil_A)$, is the set of states that can be attained by acting on the reduced state of $\ket{\TFD(t_{\rm i})}$ on $\Hil_A$ with unitary operators that correspond to the specific set of allowed $\phi_L$,
\begin{equation} \label{eq:infdomain}
D(\mathcal{N}_{A \rightarrow B}) \equiv \left\{ \rho_A = \oh_A \left( \Tr_{\bar{A}} \ketbra{\TFD(t_{\rm i})}{\TFD(t_{\rm i})} \right) \oh_A^\dagger \right\} .
\end{equation}
Here, $\oh_A$ is the CFT representation of $\phi_L$ on the boundary subregion $A$.

Such a channel is straightforward to write down and intuitive in its meaning.
It takes as input the density matrix on $A$, which describes the ingoing perturbation from the dual gravitational point of view, and outputs the reduced density matrix on $B$, which describes the perturbation that has exited the wormhole after traversal.
However, it is a channel between infinite-dimensional Hilbert spaces, to which many of the finite-dimensional results do not necessarily directly apply (see Refs.~\cite{Giovannetti,PhysRevLett.98.130501} for more discussion of infinite-dimensional bosonic channels).
Nevertheless, on one hand, we can ask how the dual gravitational description informs such channels between infinite-dimensional spaces.
On the other hand, as we will now consider, it is also interesting to try to make contact with existing results on channels between finite-dimensional spaces.

\subsection{A map between code subspaces}\label{sub:channel_code}
\begin{figure}[h!]
\centering
\includegraphics[scale=1]{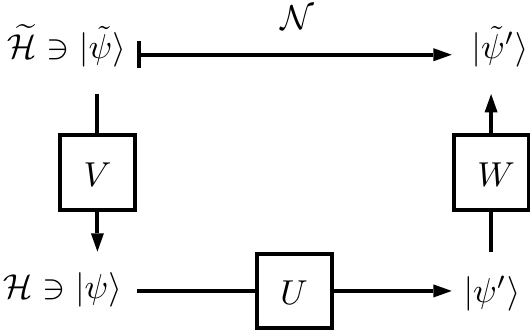}
\caption{Traversable wormhole as a map between code subspaces.}
\label{fig:wh_channel}
\end{figure}

Our aim is to construct a map that acts on states in an associated finite-dimensional Hilbert space $\widetilde \Hil$.
Furthermore, we would like $\widetilde \Hil$ to factorize as $\widetilde \Hil = \widetilde \Hil_L \otimes \widetilde \Hil_R$ in such a way that we can relate $\widetilde \Hil_L$ to excitations of the left CFT and $\widetilde \Hil_R$ to excitations the right CFT.
With these aims in mind, our strategy will be to define a map that encodes states $|\tilde \psi\rangle \in \widetilde \Hil$ as states $\ket{\psi} \in \Hil$.
We can then let $\ket{\psi}$ evolve according to the CFT unitary time evolution, including the double-trace deformation in the evolution.
Finally, by completing the procedure with a decoding of the final state back to a state in $\tilde \Hil$, the result is a mapping between states in $\tilde \Hil$.
The whole procedure is illustrated schematically in \Fig{fig:wh_channel}.

In the spirit of \Ref{Almheiri:2014lwa}, let us consider building up a collection of states that are perturbatively close to the thermofield double by acting with local bulk operators.
For simplicity, we will only consider a single type of bulk operator, $\phi(x)$, and suppose that it can be inserted at locations $x_1^L$, $x_2^L$, \dots, $x_N^L$ in the left asymptotically-AdS region and at locations $x_1^R$, $x_2^R$, \dots, $x_N^R$ in the right asymptotically-AdS region, with at most a single insertion at any location.\footnote{More generally, $\phi(x)$ could also denote smeared operators centered at $x$ that create wave packets.}
This defines a collection of $d^2 = (2^{N})^2$ states,\footnote{We will use $\ket{\TFD}$ to denote both the CFT state and the state of the dual gravitational theory, relying on context to distinguish between the two.}
\begin{equation}
\ket{\TFD(t_{\rm i})}, ~ \phi(x^L_1) \ket{\TFD(t_{\rm i})}, ~ \phi(x^R_2) \ket{\TFD(t_{\rm i})}, ~ \phi(x^L_1) \phi(x^R_2) \ket{\TFD(t_{\rm i})}, ~ \dots 
\end{equation}
This of course constitutes a coarse-graining of the full traversable wormhole picture.
We only consider a finite number of excitations at a finite number of locations because we do not want the backreaction to be strong enough to change the background geometry nonperturbatively.
In this coarse-grained regime, each $\phi(x_i)$ should be thought of as creating an excitation that we can attempt to send through the wormhole.
Transmitting a given quantum state ``through the wormhole'' will then amount to acting with the $\phi(x_i)$ in a particular correlated way.

The framework that we have built up here can be compared to, e.g., a description of the transmission of quantum information via an optical fiber.
While sending pulses of light down an optical fiber amounts to exciting the photon field in a prescribed way and then letting the field propagate, instead of working with the full set of field-theoretic degrees of freedom it is much more convenient to work with a coarse-grained picture that describes the transmission of discrete qubits.

For simplicity, let us further suppose that the locations $x_i$ are close enough to the boundary and far enough apart so that each $\phi(x_i^L)$ can be represented as a CFT operator $\oh_{A_i}$ on the left boundary with support on a minimal boundary subregion $A_i$, such that $A_i \cap A_j = \varnothing$ for $i \neq j$.
Similarly, let $\oh_{B_i}$ and $B_i$ denote such CFT operators and subregions on the right boundary corresponding to the bulk operators $\phi(x_i^R)$.
The corresponding collection of states in the CFT is of course
\begin{equation} \label{eq:collection}
\ket{\TFD(t_{\rm i})}, ~ \oh_{A_1} \ket{\TFD(t_{\rm i})}, ~ \oh_{B_2} \ket{\TFD(t_{\rm i})}, ~ \oh_{A_1} \oh_{B_2} \ket{\TFD(t_{\rm i})}, ~ \dots 
\end{equation}
We note the important caveat that it is not rigorously known whether such representations of bulk operators on minimal boundary subregions exist when the background is the thermofield double.
However, such reconstructions are possible about an empty AdS background \cite{Hamilton:2006az}, and it has been shown that global reconstructions on the full boundary are possible for AdS-Schwarzschild \cite{Kabat:2014kfa}.
For now we will proceed with the assumption above, although we could alternatively think instead of pushing the $\phi(x_i)$ all the way to the boundary, so that they are also by definition local on the boundary.
It is interesting to study finite-dimensional constructions because many results on channels apply to finite-dimensional systems; nevertheless, we include the above caveat about our construction for completeness.
Should the construction from \Eq{eq:collection} fail to hold on rigorous grounds, we would be surprised if it were impossible to design a better finite-dimensional construction in the CFT.

\subsubsection{Encoding and evolution}

Suppose that $\widetilde \Hil = \widetilde \Hil_L \otimes \widetilde \Hil_R$ is a Hilbert space of dimension $d^2$, with $\dim \widetilde \Hil_L = \dim \widetilde \Hil_R = d \equiv 2^N$, and let $\{|\tilde X_\alpha\rangle_L\}_{\alpha=1}^d$ and $\{|\tilde X_\beta\rangle_R\}_{\beta=1}^d$ be orthonormal bases for $\widetilde \Hil_L$ and $\widetilde \Hil_R$, respectively.
Given any state in $\widetilde \Hil$,
\begin{equation}
|\tilde \psi\rangle = \sum_{\alpha,\beta = 1}^d c_{\alpha \beta} |\tilde X_\alpha\rangle_L |\tilde X_\beta\rangle_R ,
\end{equation}
we can use the collection of states in \Eq{eq:collection} to encode $|\tilde \psi\rangle$ into a state in $\Hil$ by thinking of each $\tilde X_\alpha$ as one member of the power set of $\{1, 2, \dots, N\}$:
\begin{equation}
\begin{aligned}
|\tilde \psi\rangle ~ \mapsto ~ \ket{\psi} &= \frac{V|\tilde \psi\rangle}{\Vert V |\tilde \psi\rangle \Vert}\\
&= \frac{1}{\Vert V|\tilde \psi\rangle\Vert} \sum_{\alpha,\beta = 1}^d c_{\alpha \beta} \left( \bigotimes_{i \in \tilde X_\alpha} \bigotimes_{j \in \tilde X_\beta} \oh_{A_i} \oh_{B_j} \right) \ket{\TFD(t_{\rm i})}.
\end{aligned}
\end{equation}
The encoding is realized by an operator $V:\widetilde \Hil \rightarrow \Hil$,
\begin{equation} \label{eq:almostIsom}
V = \sum_{\alpha,\beta=1}^d \left( \bigotimes_{i \in \tilde X_\alpha} \bigotimes_{j \in \tilde X_\beta} \oh_{A_i} \oh_{B_j} \right) \ket{\TFD(t_{\rm i})} \langle\tilde X_\alpha|_L \langle\tilde X_\beta|_R \, .
\end{equation}
Note that $V$ is not isometric because the encoded states, $\otimes_{i \in \tilde X_\alpha} \otimes_{j \in \tilde X_\beta} \oh_{A_i} \oh_{B_j} \ket{\TFD(t_{\rm i})}$, are not orthogonal.

Given the encoded state $\ket{\psi}$, the action of the channel itself is again just the time evolution generated by the CFT Hamiltonian, which we supplement with a double-trace deformation at $t = t_{\rm w}$.
This leads to a final (encoded) state
\begin{equation}
\ket{\psi^\prime} = U \ket{\psi} ,
\end{equation}
where
\begin{equation}
U = \mathcal{U}(t_{\rm f},t_{\rm w}) e^{ih\oh_L \oh_R} \mathcal{U}(t_{\rm w},t_{\rm i}) .
\end{equation}

\subsubsection{Decoding}\label{sec:decoding}

To complete the channel, we must map the state $\ket{\psi^\prime}$ back onto a state in $\widetilde \Hil$.
To this end, we can define the mapping 
\begin{equation} \label{eq:decode}
\begin{aligned}
\ket{\psi^\prime} ~ \mapsto ~ |\tilde{\psi}^\prime\rangle &= \frac{W\ket{\psi^\prime}}{\Vert W\ket{\psi^\prime} \Vert} \\
&= \frac{1}{\Vert W\ket{\psi^\prime} \Vert}\sum_{\alpha,\beta=1}^d w_{\alpha \beta}(\psi^\prime) |\tilde X_\alpha\rangle_L |\tilde X_\beta\rangle_R \, ,
\end{aligned}
\end{equation}
where
\begin{equation}
w_{\alpha\beta}(\psi^\prime) =  \bra{\TFD(t_{\rm f})} \left( \bigotimes_{i \in \tilde X_\alpha} \bigotimes_{j \in \tilde X_\beta} \oh_{A_i^\prime}^\dagger \oh_{B_j^\prime}^\dagger \right) \ket{\psi^\prime} \, .
\end{equation}
For shorthand, we write $\ket{\TFD(t_{\rm f})}$ for $U \ket{\TFD(t_{\rm i})}$.

Let us consider this decoding in more detail.
The basic idea is that at the later time $t_{\rm f}$, we want to see whether the initial excitations made it through the wormhole to the other boundary.
If transmission through the wormhole was successful, then they should reappear as local excitations at the later time $t_{\rm f}$.
To this end, we have introduced a new set of boundary subregions, $A_i^\prime$ and $B_j^\prime$, which may be different from the original set of boundary subregions, but should be related to them as a function of, e.g., the angle of incidence of the original excitations, possible interactions among excitations in the bulk, etc.
Likewise, these new boundary subregions have associated operators $\oh_{A_i^\prime}$ and $\oh_{B_j^\prime}$, which should correspond to smearings of possible transmitted bulk excitations onto the boundary.

\begin{figure}
\centering
\includegraphics[scale=1]{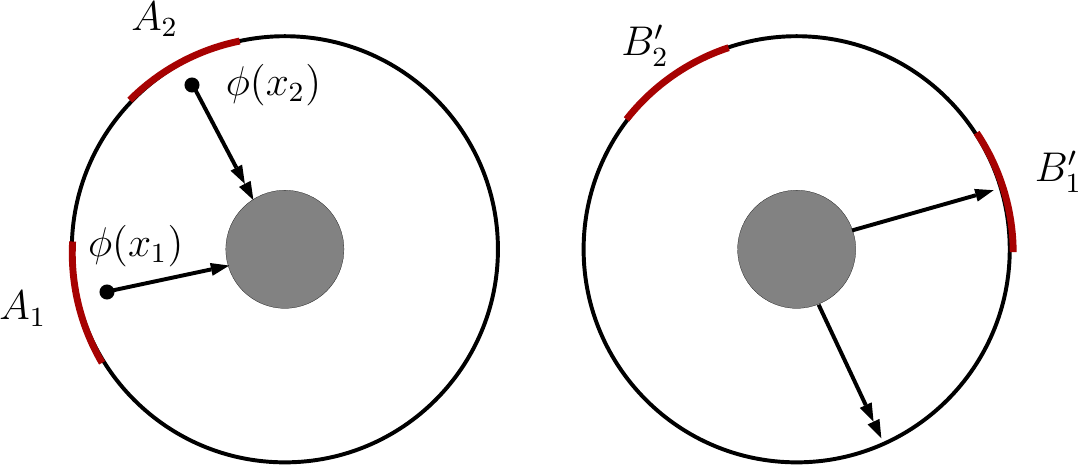}
\caption{In this example, a signal sent from the boundary subregion $A_1$ reaches the boundary subregion $B^\prime_1$, but a signal sent from $A_2$ reaches no receiving $B_j'$ for this particular choice of boundary subregions.}
\label{fig:trajectories}
\end{figure}

Different choices of $A_i$, $B_j$, $A_i^\prime$, $B_j^\prime$, and the associated operators give rise to different channels with different capacities for the same traversable wormhole.
Of course, with very poor choices of boundary subregions and operators, one could end up with channels that have artificially low capacities, as illustrated in \Fig{fig:trajectories}.
However, it seems a reasonable expectation that appropriate choices of boundary subregions and operators can adequately capture the intuitive picture of ``sending qubits through a wormhole'' with this construction.
For instance, in the limit where the excitations do not cause nonperturbative backreactions and where they do not interact in the bulk, inspection of \Fig{fig:trajectories} shows that, in a near-optimal protocol, each $B_j^\prime$ should simply be the reflection of $A_j$ in the axis perpendicular to the direction of propagation of the signal created by $\phi(x_j)$.
If the bulk excitations are allowed to interact, their propagation through the wormhole becomes a bulk scattering problem, and the $B_j^\prime$ should be chosen so as to maximize the probability of detecting transmitted excitations.
If the experimenter has sufficient resources to choose regions $B_j^\prime$ to cover the entire boundary, these considerations are unimportant, but in a resource-constrained situation (such as, for example, if the total area and/or total number of boundary subregions are limited) they become relevant.

To recapitulate, an initial state $|\tilde \psi\rangle$ is first encoded with $V$, evolved with the CFT time evolution $U$, and then decoded with an operator $W : \Hil \rightarrow \widetilde \Hil$,
\begin{equation}\label{eq:W}
W = \sum_{\alpha,\beta=1}^d |\tilde X_\alpha\rangle_L |\tilde X_\beta\rangle_R \bra{\TFD(t_{\rm f})} \left( \bigotimes_{i \in \tilde X_\alpha} \bigotimes_{j \in \tilde X_\beta} \oh_{A_i^\prime}^\dagger \oh_{B_j\prime}^\dagger \right) \, .
\end{equation}
In spirit, one can think of the map $W$ as a projection that picks out particular states in $\Hil$ that correspond to the codewords in $\widetilde \Hil$.
However, $W$ is not an exact projection, first for the simple reason that its domain and range do not coincide, so the expression ``$W^2$'' does not make sense.
Moreover, even if we consider $WV$ or $VW$, which can be repeatedly composed, one finds that $(WV)^2 \neq WV$ and $(VW)^2 \neq VW$,\footnote{In particular, the sets $A_i$ and $A_i'$ are in principle different and similarly for the $B_j$ and $B_j'$.} so neither $VW$ nor $WV$ is a projector in general.

The decoding map $W$ defined in \Eq{eq:W} has the virtue of simplicity, but it has two disadvantages. First, the normalization factor in \Eq{eq:decode} renders it nonlinear. Second, the map introduces a small amount of noise, in the sense that bulk states that correspond to excitations of $\ket{\TFD(t_{\rm f})}$ are not mapped onto single codewords.
For example, consider decoding the unexcited state $\ket{\TFD(t_{\rm f})}$ with $W$.
This state results from encoding (i.e., acting with $V$ on) the state $|\tilde \psi\rangle$ for which the only nonzero $c_{\alpha\beta}$ is the one where $\tilde X_{\alpha} = \tilde X_{\beta} = \varnothing$ (i.e., the initial state is $\ket{\TFD(t_{\rm i})}$) and then acting with $U$.
Under the action of \Eq{eq:decode}, the state $\ket{\TFD(t_{\rm f})}$ gets mapped to
\begin{equation}
\ket{\TFD(t_{\rm f})} ~ \mapsto ~ \frac{1}{C} \sum_{\alpha, \beta = 1}^d \left\langle \otimes_{i \in \tilde X_\alpha} \otimes_{j \in \tilde X_\beta} \oh_{A_i^\prime}^\dagger \oh_{B_j^\prime}^\dagger \right\rangle |\tilde X_\alpha\rangle_L |\tilde X_\beta\rangle_R \, ,
\end{equation}
where the expectation value is with respect to $\ket{\TFD(t_{\rm f})}$ and where $C$ is the required normalization.
One of the expectation values will be equal to 1, namely, the term with $\alpha$ and $\beta$ such that $\tilde X_{\alpha} = \tilde X_{\beta} = \varnothing$.
However, the other expectation values will generically be nonzero, albeit very small compared to unity provided that the boundary subregions $A_i^\prime$ and $B_j^\prime$ are small and far apart, since then the thermal expectation values will decay exponentially in the distance separating any pair of subregions on the boundary.
In the case where the $\oh_{A_i^\prime}$ and $\oh_{B_j^\prime}$ result from pushing a pointlike bulk operator $\phi(x)$ all the way to the boundary, then these other expectation values will in fact vanish.
This is because if different $\phi(x_i^\prime)$ and $\phi(x_j^\prime)$ lie on the boundary, then they are separated by infinite geodesic distance in the bulk and so their correlator vanishes. Note that this depends on the operators having low enough scaling dimensions so as to not be dual to bulk fields so massive as to have nontrivial backreaction effects close to the boundary of the spacetime.

Finally, for the same reasons, it follows that the overall map $\,\mathcal{N}$, which is the composition of encoding with $V$, evolving with $U$, and decoding with $W$, is only \emph{approximately} a bipartite unitary channel.
Because of the nonlinearity, the noisy decoding as discussed above, and additionally because the encoded codewords that result from acting with $V$ are not exactly orthogonal, the overall map $\mathcal{N}$ does not strictly describe a unitary rotation of the basis vectors $|\tilde X_\alpha\rangle_L |\tilde X_\beta\rangle_R$.
While the map remains bipartite by construction, it is not exactly unitary.
This is a further price to pay for the finite-dimensional coarse-graining.
Note, however, that exact unitarity is restored in the limit of pointlike bulk operators for light bulk fields when pushed to the boundary, as described above.

\subsubsection{Two-qubit example}

The overall map we have defined is quite abstract, so to conclude the subsection we present a simple concrete example that exhibits all of the subtleties of the encoding and decoding procedure.

Let $\widetilde \Hil = \mrm{span}\{|\tilde{0}\rangle, |\tilde{1}\rangle\}$, and consider embedding states in $\widetilde \Hil$ into the 2-qubit Hilbert space $\Hil$ according to the following linear map $V: \widetilde \Hil \rightarrow \Hil$,
\begin{equation}
\begin{aligned}
& |\tilde{0}\rangle ~ \mapsto ~ V|\tilde{0}\rangle =  \ket{0} \otimes \ket{0} \\
& |\tilde{1}\rangle ~ \mapsto ~ V|\tilde{1}\rangle = \epsilon \ket{0} \otimes \ket{0} + \sqrt{1-\epsilon^2} \ket{1} \otimes \ket{0} \, .
\end{aligned}
\end{equation}
We take the basis states $\ket{0}$ and $\ket{1}$ to correspond to spin eigenstates in the $z$ direction.
Here, $0 \leq \epsilon \leq 1$ is a parameter that controls the extent to which $V$ deviates from being an isometry (the case when $\epsilon = 0$).
Also note that this map does not preserve normalization.
We give the proper normalization below.

Let us suppose that, following encoding, the state in $\Hil$ undergoes unitary evolution according to the unitary operator $U = \sigma_x \otimes \sigma_x$. 
Then, to go back to $\widetilde \Hil$, we decode using the linear map $W : \Hil \rightarrow \widetilde \Hil$,
\begin{equation}
W = |\tilde{0}\rangle \left(\bra{1} \otimes \bra{1}\right) + |\tilde{1}\rangle \left( \epsilon \bra{1}\otimes\bra{1} + \sqrt{1-\epsilon^2} \bra{0} \otimes \bra{1} \right) \, .
\end{equation}
This lets us define an overall map $\mathcal{N} : \widetilde \Hil \rightarrow \widetilde \Hil$ where, for any $|\tilde{\psi}\rangle \in \widetilde \Hil$,
\begin{equation}
|\tilde{\psi}\rangle ~ \mapsto ~ \mathcal{N}(|\tilde{\psi}\rangle) = \frac{WUV |\tilde{\psi}\rangle}{\Vert WUV |\tilde{\psi}\rangle \Vert} \, .
\end{equation}

It is straightforward to show that a state $\tket{\tilde \psi} = \alpha \tket{\tilde 0} + \beta \tket{\tilde 1}$ gets mapped to
\begin{equation}
\mathcal{N}(\alpha \tket{\tilde 0} + \beta \tket{\tilde 1}) = \frac{1}{\left[1 + 4 \epsilon \myRe(\alpha^* \beta) + \epsilon^2 \right]^{1/2}} \left[ (\alpha + \beta \epsilon) \tket{\tilde 0} + (\alpha \epsilon + \beta) \tket{\tilde 1} \right] \, .
\end{equation}
By inspection, one can see that $\mathcal N$ is neither linear nor unitary.
However, $\mathcal{N}$ is still a positive map, as can be checked by explicit calculation:
\begin{equation}
\tbra{\tilde \psi} \mathcal{N}(|\tilde{\psi}\rangle) = \frac{1 + 2 \epsilon \myRe(\alpha^* \beta)}{1 + 4 \epsilon \myRe(\alpha^* \beta) + \epsilon^2 } \geq 0,
\end{equation}
recalling that, since $|\alpha|^2+|\beta|^2=1$, the minimum value of $\myRe(\alpha^* \beta)$ is $-1/2$.
The map $\mathcal{N}$ is strictly positive if $\epsilon < 1$.
Moreover, when $\epsilon = 0$, $\mathcal{N}$ reduces to the identity operator on $\widetilde \Hil$, which is trivially unitary.
In this trivial case, states in $\widetilde \Hil$ are orthogonally embedded in $\Hil$ with $V$ and so they can still be projectively pulled back to $\widetilde \Hil$ with $W$ following a unitary rotation by $U$ in $\Hil$.
The operators $V$, $U$, and $W$ defined here are completely analogous to the corresponding operators in the traversable wormhole setting.

\subsection{Quantum Channel Capacity}\label{sec:quantum_channel_capacity_wormhole}

For general quantum systems, including those frequently used in real-world laboratory settings, computing or bounding the quantum channel capacity is often computationally difficult or intractable \cite{Gyongyosi:2018}.
However, in the holographic setup of the traversable wormhole, additional geometric tools are at our disposal for this task. 

Strictly speaking, the map $\mathcal{N}$ is only approximately a quantum channel---it lacks linearity and complete positivity---and so it does not have a channel capacity in the definitional sense of \Sec{sec:channels}.
Nevertheless, the entanglement capacity of $\mathcal{N}$ is precisely defined, since the entanglement measures used to define entanglement capacity do not depend on the intervening map being a channel.
Since $\mathcal{N}$ is very close to being a quantum channel, it is interesting to still treat entanglement capacity as a bound on the asymptotic capacity of $\mathcal{N}$ for quantum communication.
In its unitary limit, $\mathcal{N}$ certainly has a channel capacity in a strict sense, as does the map $\mathcal{N}^{A \rightarrow B}$ defined in \Eq{eq:infchannel} for the full CFT, which is a channel by construction.

Recall from \Eq{eq:QgtrE} that the entanglement capacity $E^{(*,\text{cc})}_{c\rightarrow d,{\cal N}}$ provides a lower bound for the channel capacity $Q$.
Since an optimal $t$-shot protocol is at least as efficient as $t$ single uses of $\mathcal{N}$ for any $t$, it follows that the asymptotic entanglement capacity $E^{(*,\text{cc})}_{c\rightarrow d,{\cal N}}$ is at least as large\footnote{Since the map ${\cal N}$ associated with sending signals through the wormhole,  which we constructed in \Secs{sub:channel_subregions}{sub:channel_code}, is (to a very good approximation) a bipartite unitary, it is highly plausible that $E^{(*,\text{cc})}_{c\rightarrow d,{\cal N}} = E^{(1,*)}_{c\rightarrow d,{\cal N}}$, per \Eq{eq:oneshotequalsmultishot}. This is certainly true in the limit where $\mathcal{N}$ becomes an exact bipartite unitary map and also plausible for the gravitational reasons discussed below.} as the one-shot entanglement capacity $E^{(1,*)}_{c\rightarrow d,{\cal N}}$.

This one-shot capacity is still difficult to compute in principle.
However, on classical gravitational grounds, we can place a lower bound on the best one-shot entanglement capacity---and hence also (approximate) channel capacity--- achievable with a construction of the type described in \Sec{sub:channel_code}.
Essentially, because the traversability of the wormhole is sustained by a negative energy shock, sending signals (i.e., qubits) through the wormhole, which have positive energy, tends to make the wormhole nontraversable.
Supposing that $N_\mrm{max}$ qubits can be sent through the traversable wormhole before it becomes nontraversable, these qubits can be used to share $N_\mrm{max}$ Bell pairs between the left and right sides.
Therefore, the best one-shot entanglement capacity (resulting from the most judicious choice of boundary subregions, operators, etc.) must be at least as large as $N_\mrm{max}$.
Altogether, the bound reads
\begin{equation}
Q({\cal N}) \geq E^{(*,\text{cc})}_{c\rightarrow d,{\cal N}} \geq E^{(1,*)}_{c\rightarrow d,{\cal N}} \geq N_\mrm{max} \, .
\end{equation}
Alternatively, we can think of $N_\mrm{max}$ as providing a lower bound on the one-shot entanglement capacity $E^{(1,D)}_{c\rightarrow d,\mathcal{N}^{A \rightarrow B}}$ of the channel $\mathcal{N}^{A \rightarrow B}$ \emph{on the full CFT}, i.e., between the infinite-dimensional Hilbert spaces corresponding to boundary subregions $A$ and $B$.
The resource $D$ denotes that we only allow the preparation of states in $D(\mathcal{N}^{A \rightarrow B})$, cf. \Eq{eq:infdomain}.
Because we are not granted the resource $(*)$ here and also because the Hilbert spaces involved are infinite-dimensional, we cannot invoke the additivity results for entanglement capacities of \Ref{Bennett:2003} to obtain a similar bound on channel capacity in this latter case.

Let us now try to estimate $N_{\rm max}$, as dictated by the classical gravitational dynamics in the bulk. In this case, the optimal arrangement of signals is to group them all together into a brief packet that is sent through the wormhole at the earliest possible time. The reason for this burst-type algorithm is as follows. Following \Ref{Maldacena:2017axo}, we have a minimum bulk energy $\varepsilon$ per pulse near the horizon, from the requirement that each pulse have Compton wavelength small enough to let it fit inside the wormhole throat, which has size $\ell \Delta v =\ell \alpha \sim hG_3 e^{r_{\rm h} t_{\rm w}/\ell^2}$, so 
\be 
\varepsilon \gtrsim 1/hG_3 e^{r_{\rm h} t_{\rm w}/\ell^2}.
\ee
When the signal pulse, with positive null energy, is sent through the wormhole, it has the effect of counteracting the double-trace deformation, effectively lowering $\alpha$.\footnote{Specifically, since the signal pulse, going in the $u$ direction, must fit though the finite aperture of the wormhole throat in the $v$ direction, the uncertainty relation for $v$ implies that the pulse must carry nonzero energy-momentum in the $v$ direction, i.e., positive null energy in the same direction as the (negative null energy) shock wave that opened the wormhole in the first place. Hence, the pulse contributes its own shift in the $v$ coordinate of the horizon, partially counteracting $\alpha$. Since we are working in the shock wave approximation for the pulse as well, these effects add linearly.} Once this happens, all subsequent pulses have a smaller window in $v$ during which they can traverse the wormhole. Moreover, pulses sent at later, rather than earlier, boundary times by definition have a smaller effective wormhole window. These two effects both indicate that the information-carrying capacity of the wormhole is optimized by sending information through in a short burst of pulses. To maximize the number of pulses, let us take $\varepsilon$ to saturate this bound. Each pulse will fractionally decrease $\alpha$ by $\sim\varepsilon/|E|$, so sending too many pulses closes the wormhole entirely. This happens when the number of pulses goes as 
\be
N_{\rm max} \sim  \frac{|E|}{\varepsilon} \sim \frac{\ell E^2}{M} e^{r_{\rm h}t_{\rm w}/\ell^2} \sim \alpha \ell |E|,
\ee
using \Eqs{eq:hErelation}{eq:alpha}. We note that this value for $N_{\rm max}$ is much larger than the number of qubits computed in \Ref{Maldacena:2017axo}, since we are calculating a different quantity. Unlike \Ref{Maldacena:2017axo}, we are not requiring all of the information to be sent in the time when the probe approximation is valid. Indeed, it seems that the channel remains usable at a time during which the probe approximation is not valid---i.e., the effect of the backreaction of the qubits on the channel itself is not small---but that is also not at late times, suggesting a nontrivial channel capacity during this period. That is, by sending all of the signal at once in such a way that the wormhole is closed behind the signal, we are in effect computing the one-shot entanglement capacity of the traversable wormhole channel, in a situation where negligible backreaction is not a prerequisite.\footnote{Also note that this burst protocol is describable by the finite-dimensional formalism in \Sec{sub:channel_code}, where all of the excitations are prepared at the same initial time $t_{\rm i}$. A small change to the formalism would be necessary to describe staggered signaling, but either way, a staggered protocol is not optimal.}

Moreover, if in computing entanglement capacity we demand that the only allowed protocols are those which manifestly have a classical gravitational description, then spacetime structure implies that the entanglement capacity for multiple copies of ${\cal N}$ is additive.
Since our channel is composed of two disconnected asymptotic regions of spacetime connected by a wormhole, $N$ copies of the channel consists of $N$ pairs of asymptotic regions, each pair connected by a wormhole.
With this gravitational restriction, there is no way to compose individual uses of the channel by feeding outputs of a single channel use into a subsequent input because each channel use corresponds to a disconnected region of spacetime.
In other words, the existence of a classical gravitational description for an $N$-shot protocol means that only evolution by an $N$-fold tensor product Hamiltonian is allowed. 
Such a tensor product Hamiltonian has no capacity to generate further entanglement between the collection of boundary pairs beyond that generated between each pair individually.

We also remark that since the wormhole interiors are topologically distinct---being disconnected regions of spacetime---physical locality implies that any additional processes that take place within different wormholes during transmission must be independent and uncorrelated.
For example, one might envision refining the channel proposal by allowing bulk interactions among ingoing signals or stronger gravitational backreaction, represented via some error model.
Physical locality then implies that possible errors should be uncorrelated among channel instances.

\section{Entanglement Witnesses}\label{sec:witness}
In Section \ref{sec:observable}, we noted that although it is impossible to determine with certainty whether a wormhole connects two asymptotic regions even when the wormhole can be rendered traversable, it should nevertheless be possible to use successful signal propagation between the two regions to learn about the initial entanglement structure between the regions.
The appropriate information-theoretic tool to make this notion precise is the entanglement witness.
In quantum information theory, an entanglement witness is an operator that determines whether or not a state has a specific entanglement structure. Formally, a (partial) entanglement witness is defined as follows \cite{Horodecki:1996nc}.
\begin{defn} An operator $X$ on a bipartite Hilbert space $\Hil_A\otimes \Hil_B$ is called a \emph{(partial) entanglement witness} if there exists at least one density matrix $\rho_{AB}$ such that:
\begin{itemize}
\item[$i$.] $\rho_{AB}$ is not separable (i.e., cannot be written as $\sum_i p_i \rho_A^{(i)} \otimes\rho_B^{(i)}$) across the bipartition between $\Hil_A$ and $\Hil_B$.
\item[$ii$.] $\Tr[X\rho_{AB}]\leq 0$.
\item[$iii$.] $\Tr[X\sigma_{AB}]\geq 0$ for all separable states $\sigma_{AB}$ across the same bipartition.
\end{itemize}
\end{defn}

A perfect entanglement witness---one that, given a state of unknown entanglement between two subsystems, can determine whether that state is separable across the bipartition---cannot exist, by linearity of quantum mechanics \cite{bib:nc}. However, partial entanglement witnesses, capable only of distinguishing particular entangled states from separable states, are permitted. 

As a concrete example, let the factors $\Hil_A$ and $\Hil_B$ each describe one qubit and consider the operator $I\otimes T$, where $T$ is the transpose operator in a particular basis. We define a new operator $X$ to be $I \otimes T$ applied to the density matrix $(\ket{00}+\ket{11})(\bra{11}+\bra{00})$:
\begin{equation}
X \equiv I\otimes T (\ket{00}+\ket{11})(\bra{11}+\bra{00})=\ket{00}\bra{00}+\ket{11}\bra{11}+\ket{01}\bra{10}+\ket{10}\bra{01}.
\end{equation}
Viewed as a matrix, $X$ has an eigenvector $\ket{01}-\ket{10}$ with eigenvalue $-1$. Thus, constructing $\rho_{AB}$ in our definition from this eigenvector, $\rho_{AB} \equiv (\ket{01}-\ket{10})(\bra{01}-\bra{10})$, indeed gives a negative value of $\Tr[X\rho_{AB}]$. On the other hand, since $T$ is a positive linear map, a theorem of Peres \cite{Peres:1996dw} implies that acting with $X$ on the density matrix of any separable state yields an operator with nonnegative trace. That is, $X$ is a partial entanglement witness capable of differentiating the Bell state $\rho_{AB}$ from a separable one.

Generally speaking, the information supplied by entanglement witnesses is more detailed (but also more restricted) than the information supplied by generic measures of correlation, such as entanglement entropy.
Continuing the example above, suppose that an experimenter is supplied with many copies of an unknown pure state $\ket{\psi}$.
We may choose to expand in the basis of Bell states,
\begin{equation}
\ket{\psi} = c_1 \ket{\Phi^+} + c_2 \ket{\Phi^-} + c_3 \ket{\Psi^+} + c_4 \ket{\Psi^-},
\end{equation}
where $\ket{\Phi^{\pm}} = \tfrac{1}{\sqrt{2}} (\ket{00} \pm \ket{11})$ and $\ket{\Psi^{\pm}} = \tfrac{1}{\sqrt{2}} (\ket{01} \pm \ket{10})$.
Notice that $X$ acts trivially on the other Bell states besides $\ket{\Psi^-}$.
It consequently follows that
\begin{equation}
\Tr \left[ X \ketbra{\psi}{\psi} \right] = |c_1|^2 + |c_2|^2 + |c_3|^2 - |c_4|^2 = 1 - 2 |c_4|^2 .
\end{equation}
Therefore, in this situation, the experimenter can deduce the magnitude $|c_4|$ by measuring the expectation value of the entanglement witness $X$.
An entanglement witness reveals information about the structure of a state, which, holographically, will amount to probing the structure of wormholes that connect black holes.

Accordingly, let us consider a holographic setup. In particular, we can consider applying local unitaries on either side of the bipartition to enact gravitational collapse, converting a particular possibly-entangled pure state---for which one wants to investigate the entanglement structure---into two black holes, one made of each subsystem, without changing the entanglement structure between the two sides. In this construction, one can ask whether it is possible to construct a holographic realization of entanglement witnesses for specific patterns of entanglement.

\section{Traversable Wormholes as Entanglement Witnesses}\label{sec:TWEW}
The traversable wormhole construction allows for repeatability: one is not constrained to send a single signal as in the case of \Ref{Bao:2015nca}, but rather can send a number of signals proportional to the (negative) energy of the shock wave used to open the wormhole, as we discussed in \Sec{sec:quantum_channel_capacity_wormhole}. Thus, one is free to send multiple light pulses through the wormhole region and to ask which (and how many) successfully make it out of the other black hole. Using this freedom, one can achieve various different goals using the traversable wormhole. For example, as discussed explicitly in \Sec{sec:quantum_channel_capacity_wormhole}, one can use the traversable wormhole to send information from the left to the right side, in which case one would choose the times of the ingoing signals so as to maximize the information passing through and thereby optimize the utility of the wormhole as a quantum channel (i.e., to maximize its channel capacity). 

However, there are other uses for the traversable wormhole. In particular, one can arrange the ingoing photons in a signal sent into the wormhole in order to obtain information about the nature of the wormhole geometry itself.
That is, we can effectively implement wormhole tomography by scanning the geometry, measuring the position and time delay of signals sent through the wormhole at different times and angles of incidence. Furthermore, because the structure of the wormhole is dual to the structure of entanglement between the two black holes, the characterization of which and how many light pulses make it through the traversable wormhole serves as a set of useful partial entanglement witnesses that partially classifies the set of entangled states dual to traversable wormholes. 

As a step towards the goal of wormhole tomography, we consider the following concrete setup. For the wormhole geometry described in \Sec{sec:AdSCFT}, we have $v = e^{r_{\rm h}t/\ell^2}$ on the boundary of the right side, where both $t$ and $v$ increase toward the future. The double-trace deformation is performed on both black holes simultaneously in an attempt to create a traversable wormhole, opening up an interval in boundary time $(t_{\rm i},t_{\rm f})$ during which pulses traveling on radial geodesics, sent from the right boundary, will pass through the wormhole. In the setup shown in \Fig{fig:geometry}, a signal released from the boundary with Kruskal coordinate $v\in (0,\alpha)$ is able to traverse the wormhole; that is, $t_{\rm i} = -\infty$ and  $t_{\rm f} = \frac{\ell^2}{r_{\rm h}}\log \alpha$, with $\alpha$ given by \Eq{eq:alpha}. The goal of the experiment is to measure $\alpha$, which in this setup is unknown to the experimenter sending in the photon pulses. At some early but finite time $t_0$, a set of light pulses is sent into one of the black holes with time separation $\delta t$ between each pulse. If the experimenter wishes to measure $\alpha$ to within some given fractional precision, then in the absence of further knowledge, they would naively wish to take $\delta t$ to be as small as possible, within the engineering constraints of their apparatus, and continue to send in pulses until they cease to be received on the other side. 

However, there are competing effects between the energy of the pulses, which will backreact to close the wormhole faster for higher energy pulses, and their spacing. In order to probe constant intervals in $\alpha$ close to the wormhole, one needs to construct pulses that are exponentially close together at their source, since $\delta v = (r_{\rm h} \delta t/\ell^2)e^{r_{\rm h}t/\ell^2}$. As noted in \Sec{sec:quantum_channel_capacity_wormhole}, with each (positive null energy) signal pulse, the window in $v$ for which a signal will traverse the wormhole is decreased by $\varepsilon \alpha/|E|$. The ideal timing and energy distribution of signal pulses for the purposes of measuring $\alpha$ would depend on the experimenter's initial prior probability distribution for $\alpha$. Such considerations could impact the spacing of the pulses, the timing of the earliest pulse, and the energy of each pulse; by gaining the benefit of short, high-energy pulses, the experimenter would pay the cost of being able to send fewer of them before the wormhole closes. If the experimenter has a known bound on $\alpha$, then this sets the time at which the first or last pulse should be sent through. There is an incentive to not send unnecessary pulses, so as to prevent premature closing of the traversable wormhole. If the experimenter has access to some theoretical model-dependent prediction for $\alpha$ with some uncertainty, then the pulses should be spaced in such a way as to closely probe around this specific value, eschewing pulses that would come close to the wormhole both significantly before or after the target time. Given a prior prediction for $\alpha$, a particular experiment yields a more-precise posterior distribution according to the usual Bayesian framework; the experimentalist should design the experiment, i.e., the precise set of pulses to send into the wormhole, to maximize the information learned, i.e., minimize the entropy of the posterior distribution for $\alpha$, subject to their prior and resource limitations.

The subset of light pulses that manages to traverse the wormhole demonstrates what time window the signals must have been sent across the apparent horizon of the sender's black hole in order to traverse the wormhole, thus characterizing how traversable the wormhole was, i.e., the value of $\alpha$. This information can be used by the sender to constrain the set of unknown wormhole geometries to which the double-trace deformation could have been applied. Of course, the experimenter can really only ever directly measure $\alpha - v_0$, where $v_0$ characterizes the unknown shift in boundary time depicted in \Fig{fig:shift} and discussed in \Sec{sec:observable}.\footnote{In the notation of \Sec{sec:observable}, $v_0$ would equal $e^{r_{\rm h}(t-t_0)/\ell^2}$, where the state $\ket{\psi(t_0)}$ at some reference boundary time $t_0$ is  $\ket{\phi(t)}$ for some value $t$ among the states given in \Eq{eq:states_t_shift}.}  (Equivalently, the experimenter measures a combination of $\alpha$ and the boundary time at which the double-trace deformation was turned on.) This information, in turn, can be used via the AdS/CFT correspondence to constrain the subset of entangled states that the two black hole system could have been in, thus constructing a holographic dual of a set of entanglement witnesses as discussed in \Secs{sec:observable}{sec:witness}.

It should be noted that, as constructed here, each light pulse is, by itself, an entanglement witness: it will never reach the other side for a product state of the two black holes, and it will reach the other side for some subset of entangled states. By repetition of this process, one can gain a great deal of information constraining the kind of entangled state that the two black hole system is in. In particular, the entangled states dual to wormholes can be classified by the length of the wormhole in the dual picture, introducing perhaps an alternative information-theoretic notion to the complexity that would grow with the length of the wormhole \cite{Brown:2015bva}.

\section{Discussion}\label{sec:discussion}

\begin{emergency}{5pt}
In this paper, we have analyzed the double-trace deformation that renders wormholes traversable from a quantum information-theoretic perspective.
We have argued that the process of sending signals from one asymptotic bulk to another through the wormhole is best thought of as a quantum channel and that the ability to send multiple such signals allows the experimenter to learn about the state of the wormhole. 
\end{emergency}

Although our analysis builds on the detailed constructions within the AdS/CFT framework of Refs.~\cite{Gao:2016bin, Maldacena:2017axo}, our conclusions are largely independent of these details: all we require is that the experimenter on the boundary have access to some operation that sources the negative averaged null energy necessary to make a wormhole traversable. 
In particular, although the double-trace deformation in the CFT description creates an excitation that is manifestly entangled between the two sides, the gravitational construction in \Sec{sec:AdSCFT} does not require this. 
It is only necessary that each shock wave carry negative averaged null energy; in fact, in the limit that backreaction is small, we can treat the two shock waves as entirely independent. 

We have described the traversable wormhole in \Sec{sub:channel_subregions} as a quantum channel that maps an excitation localized near one boundary subregion to an excitation on the other boundary.
In the CFT, such a channel should be relatively straightforward to construct: the evolution map, as constructed schematically in \Eq{eq:fullUnitary}, should be built only from normal boundary time evolution, the double-trace deformation coupling the two boundaries, and the insertion of the source at the boundary.
We could imagine building up the state using the Euclidean path-integral construction on a Riemann surface formed by two thermal cylinders linked by the deformation.
The statement that the double-trace deformation renders the wormhole traversable means that the effect of the channel is simply to transfer excitations from one boundary to the other (with appropriate redshift factors, etc.), which implies a relation between time evolution and the deformation itself.
This question has recently been investigated \cite{Kourkoulou:2017zaj,Maldacena:2018lmt} in the context of AdS${}_2$ gravity and the SYK model, as well as in explicit four-dimensional constructions \cite{Maldacena:2018gjk}. It would be interesting to pursue it in a more general CFT context.

\begin{emergency}{5pt}
Furthermore, we have characterized the quantum channel corresponding to passage through the wormhole as a map between (finite-dimensional) code subspaces (\Sec{sub:channel_code}). 
One of the lessons of our approach, compared to the initial discussion of Refs.~\cite{Gao:2016bin, Maldacena:2017axo}, is that it is more natural to think of the propagation of excitations from one boundary to the other not in terms of quantum teleportation but instead as the direct, physical movement of excitations from one boundary to the other through the bulk geometry that includes the traversable wormhole. 
This picture has interesting implications on the entanglement structure of the theory, as well as lessons for how classical bulk geometries are encoded in the CFT, which have recently been discussed \cite{jafferis}.
\end{emergency}

Subsequently, in \Sec{sec:quantum_channel_capacity_wormhole} we used the gravitational dual description of the double-trace-deformed thermofield double state to bound the entanglement capacity (and hence the quantum channel capacity) of the quantum channel describing the deformation. We found that the existence of a holographic description of the state as a traversable wormhole makes the calculation of this capacity bound tractable. Specifically, we defined a protocol in the bulk that can be used to maximize the number of qubits that can be sent through the traversable wormhole.  It would be interesting to consider what other information-theoretic quantities for holographic states can be computed gravitationally and whether the channel capacity could be computed for holographic states other than the single traversable wormhole.

Finally, in \Sec{sec:TWEW} we considered the traversable wormhole as an entanglement witness. If an experimenter has access to the traversable wormhole, but does not know the magnitude of the double-trace deformation---that is, does not know ``how open'' the wormhole is, as defined by the parameter $\alpha$---then they can try to measure this quantity by sending signals into the wormhole and checking for which signals make it through. In doing so, the experimenter measures a combination of $\alpha$ and the time that the double-trace deformation was applied. We discussed the optimal protocol for making this measurement; its interpretation as an entanglement witness follows from \Secs{sec:observable}{sec:witness}.

More broadly, in this paper we have presented a quantum information-theoretical description of the bulk containing a wormhole.
The presence of the wormhole can be recognized by the fact that excitations sent from one side through the channel corresponding to the bulk geometry arrive on the other side (relatively) undisturbed, having propagated through mostly empty space.
It is tempting to conjecture that this picture applies more broadly to give a general quantum information-theoretic definition of holography.
That is, in general, if we have some strongly-coupled theory it is an extremely difficult field-theoretic problem to determine when a dual classical bulk description exists.
However, we seem to have found a simple criterion in the language of quantum channels: such a bulk description exists when there exists a channel that translates localized excitations from one portion of the theory to the other in a controlled way, corresponding to the dual of the excitation traveling through the bulk from one part of the boundary to the other.
It would be interesting to see if this criterion could be made more precise.

\begin{center} 
 {\bf Acknowledgments}
 \end{center}
 \noindent 
We thank Matt Hodel, Junyu Liu, Alex May, Arvin Shahbazi-Moghaddam, Douglas Stanford, Mark Van~Raamsdonk, and Quntao Zhuang for useful discussions.
N.B. is supported by the National Science Foundation under grant number 82248-13067-44-PHPXH.
A.C.-D. was supported by a Beatrice and Sai-Wai Fu Graduate Fellowship 
in Physics and by the Gordon and Betty Moore Foundation through Grant 
776 to the Caltech Moore Center for Theoretical Cosmology and Physics 
for the entirety of the work except for the final stages of editorial 
review. A.C.-D. is currently supported in part by the KU Leuven C1 grant 
ZKD1118 C16/16/005, the National Science Foundation of Belgium (FWO) 
grant G.001.12 Odysseus, and by the European Research Council grant no. 
ERC-2013-CoG 616732 HoloQosmos.
J.P. is supported in part by the Simons Foundation and in part by the Natural Sciences and Engineering Research Council of Canada. 
G.N.R. is supported by the Miller Institute for Basic Research in Science at the University of California, Berkeley.

\bibliographystyle{utphys-modified}
\bibliography{wormholes-submission}

\end{document}